\documentclass[manuscript]{acmart}
\renewcommand\footnotetextcopyrightpermission[1]{}
\usepackage{xspace}
\usepackage{graphicx} % added
\usepackage{multirow} % added
\usepackage{listings}
\usepackage[linesnumbered, ruled, vlined]{algorithm2e} %NOT ACCEPTED
\usepackage{lscape, float} 
\usepackage{subcaption}

\AtBeginDocument{%
  }

\begin{document}

% \newcounter{RQCounter}
% \newcommand{\hr}[1]{\textbf{RQ}$_{\ref{#1}}$}

%%
%% The "title" command has an optional parameter,
%% allowing the author to define a "short title" to be used in page headers.
%\title{How Are Hackathons Creative?}
\title{How Do Hackathons Foster Creativity? Towards AI Collaborative Evaluation of Creativity at Scale}
% \title{How Do Hackathons Foster Creativity? Towards AI Enhanced Evaluation of Creativity at Scale}

%%
%% The "author" command and its associated commands are used to define
%% the authors and their affiliations.
%% Of note is the shared affiliation of the first two authors, and the
%% "authornote" and "authornotemark" commands
%% used to denote shared contribution to the research.
\author{Jeanette Falk}
\authornote{Both authors contributed equally to this paper}
\email{jfo@cs.aau.dk}
\orcid{0000-0001-7278-9344}
\affiliation{%
  \institution{Aalborg University, Copenhagen, Department of Computer Science}
  \streetaddress{A. C. Meyers Vænge 15}
  \city{Copenhagen}
  \country{Denmark}}

  \author{Yiyi Chen}
\authornotemark[1]
\email{yiyic@cs.aau.dk}
\orcid{0000-0001-9977-8960}
\affiliation{%
  \institution{Aalborg University, Copenhagen, Department of Computer Science}
  \streetaddress{A. C. Meyers Vænge 15}
  \city{Copenhagen}
  \country{Denmark}}

  \author{Janet Rafner}
\email{janetrafner@mgmt.au.dk}
\orcid{0000-0001-9264-3334}
\affiliation{%
  \institution{Center for Hybrid Intelligence, Department of Management}
  \institution{Aarhus Institute for Advanced Studies, Aarhus University}
  \streetaddress{}
  \city{Aarhus}
  \country{Denmark}}

  \author{Mike Zhang}
\email{jjz@cs.aau.dk}
\orcid{0000-0003-1218-5201}
\affiliation{%
  \institution{Aalborg University, Copenhagen, Department of Computer Science}
    \institution{Pioneer Centre for Artificial Intelligence}
  \streetaddress{A. C. Meyers Vænge 15}
  \city{Copenhagen}
  \country{Denmark}}

  \author{Johannes Bjerva}
\email{jbjerva@cs.aau.dk}
\orcid{0000-0002-9512-0739}
\affiliation{%
  \institution{Aalborg University, Copenhagen, Department of Computer Science}
  \streetaddress{A. C. Meyers Vænge 15}
  \city{Copenhagen}
  \country{Denmark}}

  \author{Alexander Nolte}
\email{a.u.nolte@tue.nl}
\orcid{0000-0003-1255-824X}
\affiliation{%
  \institution{Software Engineering and Technology, Eindhoven University of Technology}
  \streetaddress{}
  \city{Eindhoven}
  \country{Netherlands}}
\affiliation{%
  \institution{Software and Societal Systems, Carnegie Mellon University}
  \streetaddress{}
  \city{Pittsburgh, PA}
  \country{USA}}

%%
%% By default, the full list of authors will be used in the page
%% headers. Often, this list is too long, and will overlap
%% other information printed in the page headers. This command allows
%% the author to define a more concise list
%% of authors' names for this purpose.
\renewcommand{\shortauthors}{Falk and Chen et al.}

%%
%% The abstract is a short summary of the work to be presented in the
%% article.
\begin{abstract}
Hackathons have become popular collaborative events for accelerating the development of creative ideas and prototypes.
There are several case studies showcasing creative outcomes across domains such as industry, education, and research.
However, there are no large-scale studies on creativity in hackathons which can advance theory on how hackathon formats lead to creative outcomes.
We conducted a computational analysis of 193,353 hackathon projects. By operationalizing creativity through usefulness and novelty, we refined our dataset to 10,363 projects, allowing us to analyze how participant characteristics, collaboration patterns, and hackathon setups influence the development of creative projects.
The contribution of our paper is twofold: We identified means for organizers to foster creativity in hackathons.
We also explore the use of large language models (LLMs) to augment the evaluation of creative outcomes and discuss challenges and opportunities of doing this, which has implications for creativity research at large.
\end{abstract}

%%
%% The code below is generated by the tool at http://dl.acm.org/ccs.cfm.
%% Please copy and paste the code instead of the example below.
%%
%\begin{CCSXML}
%<ccs2012>
%  <concept>
%    <concept_id>10003120.10003121.10011748</concept_id>
%    <concept_desc>Human-centered computing~Empirical studies in HCI</concept_desc>
%    <concept_significance>500</concept_significance>
%  </concept>
%  <concept>
%    <concept_id>10003120.10003130.10003131.10003235</concept_id>
%    <concept_desc>Human-centered computing~Collaborative content creation</concept_desc>
%    <concept_significance>500</concept_significance>
%  </concept>
%  <concept>
%    <concept_id>10010147.10010257</concept_id>
%    <concept_desc>Computing methodologies~Machine learning</concept_desc>
%    <concept_significance>300</concept_significance>
%  </concept>
%</ccs2012>
%\end{CCSXML}

%\ccsdesc[500]{Human-centered computing~Empirical studies in HCI}
%\ccsdesc[500]{Human-centered computing~Collaborative content creation}
%\ccsdesc[300]{Computing methodologies~Machine learning}

%%
%% Keywords. The author(s) should pick words that accurately describe
%% the work being presented. Separate the keywords with commas.
%% A "teaser" image appears between the author and affiliation
%% information and the body of the document, and typically spans the
%% page.

%\received{20 February 2007}
%\received[revised]{12 March 2009}
%\received[accepted]{5 June 2009}

%%
%% This command processes the author and affiliation and title
%% information and builds the first part of the formatted document.
\maketitle

\section{Introduction}
Reacting \textit{fast} to rapid changes in society and technology development has generally been considered a prerequisite for innovation \cite{ringel2015rising}.
Originating in niche software development environments in Silicon Valley during the late 1990's, \textit{hackathons} have become a popular format for accelerating people's creativity and developing creative ideas, prototypes, products, or services in a variety of contexts such as education \cite{schulten2024we}, entrepreneurship \cite{irani2015hackathons}, corporations \cite{pe2022corporate}, scientific communities \cite{huppenkothen2018hack}, civic engagement \cite{schrock2016civic} and others \cite{falk2020yearsOfResearch}. 
Hackathons are time-bounded, participant-driven design events, often spanning only a few days, where participants form teams and collaborate on projects to address a common theme or challenge, resulting in a perceptible outcome, such as an interactive prototype \cite{falk2024future}.
They are often celebrated for their potential for creative outcomes, see e.g.~\cite{attalah2023whoCapturesValueFromHackathons, Briscoe2014DigitalIT, taylor2018everybodyshacking, frey2016innovationdrivenHackathon, lobbe2021innovation}.
For this reason, hackathons have become an attractive approach for entrepreneurs, companies, educators and researchers to develop creative solutions to problems they face or to complement or enhance their existing innovation processes \cite{falk2020yearsOfResearch}.

Since 2016, the CHI community has shown a growing interest in hackathons. 
A search for ``hackathon'' in the CHI proceedings on the ACM Digital Library returns 87 results for full research papers.
Most research focusing on creativity in hackathons, however, consists of case studies of one or only a few events \cite{gama2023developers,rosell2014internalhackathons,Briscoe2014DigitalIT,frey2016innovationdrivenHackathon}.
While they can provide valuable insights, case studies are not suitable to advance and validate theory, either on hackathons as a phenomenon in itself or on \textit{how} hackathon formats lead to creative outcomes.
Answering fundamental research questions in this frame requires larger-scale quantitative studies in order to test hypotheses \cite{edmondson2007methodological, falk2024future}.
Current quantitative studies of hackathons focus mainly on understanding project continuation \cite{nolte2020WhatHappensToAllTheseHackathonProjects,mcintosh2021hackathonschangetheworld} and re-use of code that was created during hackathons \cite{Mahmoud2022, imam2021secretlifeofhackathoncode, imam2021trackinghackathoncode}.
To the best of our knowledge, as of yet, no research has conducted large-scale analysis of hackathon projects from a creativity perspective.
Our work addresses this gap by asking the following first research question:

\textbf{RQ1}: How can we define and implement an analysis of creativity in a way that enables large-scale analysis of creativity within hackathon projects?

Large-scale creativity assessment is essential for validating the effectiveness of the different contexts in which hackathons are organized for supporting creative outcomes -- such as in educational programs, workplace training, and other interventions -- ensuring that the formats are successfully cultivating the creative skills demanded \cite{Rafner2022DigitalGames}.
From this perspective, Rafner and colleagues have suggested that exploring methods which combine scalability ``and product-oriented assessments (data with high ecological validity) could greatly enhance the construct validity of creativity assessment instruments'' \cite{Rafner2022DigitalGames}.
In our paper, we contribute to exploring such scalable methods with potentially high ecological validity for assessing creative products in the context of hackathons, since the data we analyze resemble real-world data rather than data created in controlled creativity test settings \cite{Rafner2023creativityintheageofgenAI}.
We thus also address the following second research question:

\textbf{RQ2}: What insights into participants, collaboration patterns, and hackathon setups can we gain from analyzing a large number of creative hackathon projects and how do these insights relate to fostering creativity?

Similar to the contributions by \cite{nolte2020WhatHappensToAllTheseHackathonProjects, imam2021secretlifeofhackathoncode, imam2021trackinghackathoncode, Mahmoud2022}, we use Devpost's database of hackathons and hackathon projects for our large-scale analysis of creativity.\footnote{\url{https://devpost.com/}}
Similar to Fang, Herbsleb and Vasilescu \cite{fang2024novelty}, we operationalize a definition of creativity which we explore as a lens to analyze our dataset.
Specifically, we operationalize the concepts of \textit{novelty} and \textit{usefulness} to create a subset of creative projects.
We conduct statistical analysis on this subset of creative projects to identify patterns in the interaction of aptitude, process and environment, a contribution which we frame as particularly valuable for researchers and practitioners who organize hackathons and wish to support hackathon participants' creativity and increase the potential for creative outcomes. 

Furthermore, to address calls for exploring a Generative AI-augmented expert assessment \cite{beghetto2019large-scaleassessments, Rafner2023creativityintheageofgenAI, luchini2023automatic}, we explore the compelling case of Large Language Models (LLMs)-as-a-judge to augment our large-scale creativity evaluation method and address the following research question:

\textbf{RQ3}:How might LLMs be used to augment large-scale evaluations of creative hackathon projects?

The contribution of this two-pronged large-scale exploration of creativity in hackathon projects is the following:
In addressing RQ1, we arrive at an operationalization of creativity theory which guides the first prong of our two-pronged approach; a statistical analysis of a dataset consisting of 10,363 hackathon projects.
We discuss the findings from this statistical analysis as take-aways for organizers, thereby addressing RQ2.
As the second prong of the two-pronged approach, we address RQ3 by using LLMs to analyze the creativity of a subset consisting of 21,318 randomly sampled hackathon projects which contained both creative and non-creative projects, as defined by our operationalization (RQ1).
We also compared the ratings of LLM judges and human judges on a randomly selected subset of 30 hackathon projects. 
Drawing from the LLM analysis, we explore how LLMs can function as supplementary judges, supporting human judgment within a human-AI hybrid intelligent system \cite{Sherson2024}, and how they can facilitate large-scale analyses of hackathon creativity.
The large-scale exploration of creativity in hackathons contributes towards a greater understanding of how creativity manifests in hackathons, which we argue contributes to moving large-scale creativity evaluation towards real-world relevant scenarios with high validity. 
Finally, we discuss the challenges, limitations, and opportunities of our exploratory methods of measuring creativity on a large scale which contribute towards aiding researchers interested in quantitative instruments for measuring creativity in general.

\section{Related Work}
In this section we outline related works researching creativity in hackathons and how creativity can be evaluated. 

\subsection{Creativity in Hackathons}
Hackathons are often highlighted for accelerating participants' creativity and resulting in creative outcomes \cite{Briscoe2014DigitalIT}.
Employee-focused internal hackathons have been highlighted as a means for testing  ``new products and services as well as to generate new ideas'' \cite{rosell2014internalhackathons} and streamline creativity by bypassing slow decision-making processes and fixed organizational structures \cite{frey2016innovationdrivenHackathon}.
Not surprisingly, creative industries such as media, arts, and culture fields have also adopted and adapted hackathons \cite{karlsen2017youcandanceyourprototype}.
Karlsen and Løvlie explains this with these industries' ``long tradition of using constraints and games to facilitate creativity in arts'' \cite{karlsen2017youcandanceyourprototype}.

Previous work has explored how to organize hackathons in order to facilitate creativity.
%Based on interviews, 
%with participants from eight teams in a single hackathon, 
Ten{\'o}rio and colleagues 
%discussed some initiatives organizers may consider if they wish to support their participants' creativity.
%Their 
provide suggestions which mainly focus on pre-events such as ``workshops, coaching, training, mentoring, and so on'' \cite{Tenorio2019RethinkingSO} for: (1) supporting \textit{knowledge application}, where the participants apply and combine their individual domain knowledge \cite{olesen2018fourfactors} into new perspectives and solutions for challenges; (2) learning about how to \textit{manage conflicts}; (3) supporting participants' \textit{individual learning}; (4) organizing shorter events to mitigate fatigue \cite{Tenorio2019RethinkingSO}.
However, these recommendations are based on a small sample, making it difficult to assess their validity and understand to what extent they might hold in different contexts. 
Lobbe, Bazzaro, and Sagot researched collaborative design tools used to enhance creativity and innovation in a hackathon attended by 1,310 engineering students, resulting in around 160 projects \cite{lobbe2021innovation}.
%While analyzing a larger dataset of 160 hackathon projects, t
This larger-scale study focuses on a narrow perspective on how to support creativity contributing mainly with insights on some select tools. 
In addition, the study focused on a single event with a specific setup and population of participants. 
%We aim to gain insights that might translate to different contexts.
Broadening the scope somewhat further in terms of number of hackathons studied, Attalah, Nylund and Brem studied three different hackathons through participant observation to understand ``the impact of open innovation and collective intelligence in hackathons'' \cite{attalah2023whoCapturesValueFromHackathons}, where they frame the collective creativity taking place in hackathons as playing a key role for fostering collective intelligence. 
Collective intelligence is a kind of distributed intelligence constantly developed and coordinated by a group of people, for example, a community or society, while collective creativity is a ``momentary, collective process that includes interaction in the form of help seeking, help giving, reflective reframing and reinforcing'' \cite{hargadon2006collections} as cited in \cite{attalah2023whoCapturesValueFromHackathons}.
A key finding is that organizers should frame hackathons as a lasting development of collective intelligence, rather than momentary events of collective creativity \cite{attalah2023whoCapturesValueFromHackathons}.

These contributions are valuable in their own right to understand creativity in the context of hackathons; however, these cases take place in specific settings, which means that the findings may not be translated to other settings, as hackathons are very diverse \cite{falk2021whatdo}.
In order to advance and validate theories on creativity in hackathons, we need quantitative methods \cite{edmondson2007methodological} and large-scale analyses \cite{Rafner2022DigitalGames} in addition to case studies.

Moving from small-scale and qualitative data collections to larger ones, a couple of literature reviews have aimed at consolidating knowledge on how to organize hackathons including how to support creativity.
Kollwitz and Dinter's contribution is a taxonomy of hackathons where they ``figured out which dimensions and objectives are discussed in the literature'' \cite{kollwitz2019whatthehack} and which ``contributes to a better understanding of the opportunities and characteristics of hackathons'' \cite{kollwitz2019whatthehack}.
While they mention innovation and creativity as characteristic of hackathons, they mainly discuss their taxonomy as a way to reduce uncertainties for organizers regarding results, processes and resources, for example by making ``detailed specifications regarding the solution space as well as the degree of elaboration in order to channel the creativity of the participants in a desired direction'' \cite{kollwitz2019whatthehack}.
Furthermore, their taxonomy ``need[s] further evidence to show that those aspects are actually relevant from a practical point of view'' \cite{kollwitz2019whatthehack}.

Heller and colleagues conducted a literature review of 87 articles on how to best execute hackathons including elements which are meant to facilitate creativity \cite{Heller2023hackyourorganizationalinnovation}.
The following elements were specifically highlighted for their relation to increased creativity: 
Pre-registered teams may improve teamwork but may jeopardize creativity as they are usually made up of participants with similar backgrounds \cite{nolte2020organize} whereas diverse teams have shown increased creativity \cite{carmit2012beyondindividualcreativity}.
Heller and colleagues further found that ``a very competitive atmosphere with high-value prizes creates extrinsic motivation to stand out, but an atmosphere of collaboration leads to improved creativity, better teamwork, and intrinsic motivation'' \cite{Heller2023hackyourorganizationalinnovation}.
Drawing especially on the findings by Lifshitz-Assaf and colleagues, Heller and colleagues emphasized that although research has discussed the detrimental effect of time pressure on creativity (see e.g. \cite{amabile2002creativity}), the limited time-frame of hackathons are by design, and instead of following traditional coordination strategies participants should adopt \textit{adaptive coordination processes} \cite{Lifshitz-Assaf2021minimalandadaptivecoordination}, similar to Edmondson's distinction between teamwork and \textit{teaming} \cite{edmondson2012teaming}.
For supporting ideation, Heller and colleagues identified having mentors as ideation facilitators for increasing creativity and referred to Wilson's description of good brainstorming facilitators \cite{wilson2013brainstorming}: ``preventing participants from offering premature criticism, encouraging the flow of ideas, focusing on quantity rather than the quality of ideas, and promoting tolerance for radical ideas'' \cite{Heller2023hackyourorganizationalinnovation}.

However, similar to related works involving case studies of hackathons, Heller and colleagues' literature review included ``existing studies [that] are limited to mainly correlative case studies, [and] which do not allow for a proper understanding of the causal processes underlying effective hackathon execution'' \cite{Heller2023hackyourorganizationalinnovation}.
They further recommend future research exploring hackathons to conduct ``advanced qualitative (such as collecting previous hackathons’ post-event surveys and interviews) and quantitative methods (such as meta-analysis) to offer data-driven conclusions on how to best plan and execute hackathons'' \cite{Heller2023hackyourorganizationalinnovation}.
Our contribution falls into the latter type of research.
Heller and colleagues also suggest that ``studies can examine participant-, organization-, or event-level variables of interest by comparing the outcomes of two hackathons that are identical in all characteristics except one (e.g., staff diversity, judges’ identity, virtual versus physical versus hybrid, etc.)'' \cite{Heller2023hackyourorganizationalinnovation}.
This is, to some extent, what we contribute in this paper: However, instead of comparing only two hackathons where all except one variable are controlled, we utilized a large-scale dataset consisting of almost 200,000 hackathon projects. 
% and identified by our criteria as either creative or less creative. 

\subsection{Evaluating Creativity}
The need for accurate and valid methods to evaluate creativity was first emphasized during Guilford’s 1950 presidential address to the American Psychological Association (APA) and has been a popular topic of study in many disciplines, including the CHI community \cite{wallace2020sketchy}. 
More recently, the 2019 special issue of Philosophy, Aesthetics, Creativity and the Arts, featured an editorial statement stressing that ``without proper instruments to measure creativity or adequate standards of assessment, the validity of any creativity study is seriously questioned''  \cite{Barbot2019}. 
However, developing and adhering to rigorous creativity assessments standards proves exceptionally difficult, given the complex, contextual nature of creativity and the many ways in which it can be evaluated. 

In psychology, creativity assessments include standardized instruments targeting components of divergent thinking (e.g., originality, flexibility, fluency), as well as expert product evaluations and measures of creative behavior or self-efficacy \cite{Rafner2022DigitalGames}. One of the major drawbacks of many creativity assessments is that they require extensive human effort for manual evaluation which is labor intensive and often suffers from disagreement among raters \cite{Rafner2022DigitalGames}. 
Some automated creativity scoring leverages advances in Natural Language Processing (NLP) to address these limitations \cite{barbot2019measuring, forthmann2017missing, reiter2019scoring}.

\paragraph{LLM-as-a-judge.} Recent NLP research trends highlight a shift towards using LLMs for automated evaluation of text documents, potentially supplementing or replacing human judgments~\cite{chiang-lee-2023-large, wang-etal-2023-chatgpt, liu-etal-2023-g, zheng2024judging}. 
This approach manifests in competition leaderboards such as~\cite{zheng2024judging} or \cite{lambert2024rewardbench}, where researchers evaluate models against each other based on human evaluations or pre-defined benchmarks. 
In general, rating long and detailed survey responses based on pre-designed criteria is challenging even to human experts.
To reduce manual effort, researchers can instruct an LLM to rate any given text on, e.g., a Likert-like scale (1-7) or other custom rubrics as, e.g., proposed by Kim and colleagues~\cite{kim2024prometheus}. 
This ``LLM-as-a-judge'' approach has demonstrated strong correlations with human judgments on various natural language tasks~\cite{liu-etal-2023-g,zheng2024judging,chen-etal-2023-exploring-use,tornberg2023chatgpt,naismith-etal-2023-automated,gilardi2023chatgpt,kocmi-federmann-2023-large,huang-etal-2024-chatgpt} and an even better agreement when used in a ``jury setting'' with multiple models judging the same input~\cite{verga2024replacing} and aggregating the ratings. 

Automated methods for creativity assessment have mostly been applied to scoring divergent thinking tasks, such as the Alternative Uses Task\cite{HADAS2024101549, ORGANISCIAK2023101356}. These automated methods have focused exclusively on scoring the originality of ideas \cite{acar2023creativity, beaty2022semantic, beaty2021automating, heinen2018semantic, patterson2023multilingual} .
However, assessment of  solutions from real-world creative problem-solving, often requires both evaluating novelty and usefulness (i.e., plausible and effective) \cite{reiter2019scoring}.  

Luchini and colleagues \cite{luchini2023automatic} have taken an important step towards addressing this issue by using LLMs to automatically evaluating standard creative problem solving tasks for both novelty and usefulness.
While more complex than data from the Alternative Uses Task, their creative problem solving tasks still result in well structured, reasonably short, homogeneous data.  
There is a need for methods to handle the variability of real-world data, ensuring LLM evaluations of creativity remain as effective as in controlled settings.  

For complex and context-rich datasets such as hackathon outputs, we propose that instead of replacing subjective human evaluations, LLM technologies should assist human judges by providing scalability, analytical insights, and potentially consistency, to enhance human expertise. 
This approach aligns with recent advancements in hybrid intelligence and human-centered AI, which emphasize synergetic human-AI interactions \cite{Rafner2023creativityintheageofgenAI, Mao2023, Mao2024, Sherson2024}, fostering collaborative partnerships where AI serves as a supportive tool for humans \cite{li2018make, shneiderman2020human}.

 %By researching hackathons from a creativity perspective, we can explore outcomes which are considered creative in many ways. 
Our contribution approaches large-scale analysis of our context-rich hackathon project dataset from two angles: First, we explore how to operationalize creativity in order to conduct statistical analysis, and secondly, we explore how LLMs may be used as an additional supporting judge.
%is a bottom-up, large-scale analysis which investigates settings which may lead to creative hackathon projects,using as LLMs as an additional supporting judge. 

\section{Method}
Similar to our motivation for exploring large-scale evaluation of creativity, Fang, Herbsleb and Vasilescu explored creativity and innovation by analyzing 70,891 projects from the World of Code dataset.
They follow the Schumpeterian tradition \cite{schumpeter1939business} of ``viewing innovation as emerging from the novel recombination of existing bits of knowledge'' \cite{fang2024novelty} and analyze their data by operationalizing this at code level, by focusing on unusual combinations of software packages.
From this lens, they found that innovative projects tended to have more GitHub star counts -- in other words, novelty begets popularity.
We expand on this approach by (1) considering the established standard definition of creativity as defined by Plucker, Beghetto and Dow \cite{plucker2004standarddefinitioncreativity}, and (2) evaluate this in the context of a new dataset based on a large sample of hackathon projects (further detailed in Section~\ref{sec:setup}). 

\subsection{Creativity Framework}
In order to answer the three research questions stated in the introduction, 
%analyze our dataset,
we constructed a framework to guide our analysis and frame the results. 
This was done iteratively, by continuously exploring our dataset, familiarizing ourselves with it, reviewing related creativity literature, and discussing among ourselves about how creativity is measured in general and how we might analyze hackathon creativity based on our available data.
During these discussions, we constructed the framework which was used to scaffold the final analysis of the dataset.
To create the framework for our analysis, we took point of departure in Plucker, Beghetto and Dow's standard definition of creativity, which is based on a cross-disciplinary review of how peer-reviewed business, education, psychology, and creativity journal articles evaluated the term creativity:
\begin{quote}
``Creativity is the \textbf{interaction} among \textbf{aptitude}, \textbf{process}, and \textbf{environment} by which an individual or group produces a \textbf{perceptible product} that is both \textbf{novel} and \textbf{useful} as defined within a social context'' \cite{plucker2004standarddefinitioncreativity}
\end{quote}

\subsubsection{Operationalizing the Standard Definition of Creativity}
We approach the definition in a reverse order: 
First, we defined how \textit{novelty} and \textit{usefulness} may be expressed within the context of the dataset consisting of hackathon projects.
While there have been many definitions of creativity in addition to Plucker, Beghetto and Dow's \cite{plucker2004standarddefinitioncreativity}, this twin criteria of novelty and usefulness have formed ``principal components of numerous definitions of creativity dating back at least 70 years'' \cite{lloydcox2022evaluatingcreativity}.
This means that for a project to be creative it should be both novel \textit{and} useful, and we thereby expand on the approach taken by Fang, Herbsleb and Vasilescu \cite{fang2024novelty} who only focused on novelty.
The precise terminology of novelty and usefulness can vary \cite{lloydcox2022evaluatingcreativity}, and in our definitions of them used for the LLM-as-a-judge approach, we aimed at including oft-repeated elements from the creativity research literature. 

\paragraph{Novelty} 
Related to our exploration of LLM-as-a-judge, we took inspiration from Luchini and colleagues' prompt design, where they explored automatic scoring of Creative Problem-Solving with LLMs.
In their prompt, novelty refers ``to how unique the approach is compared to typical solutions'' \cite{luchini2023automatic}.
This resembles the strategy taken by Fang, Herbsleb and Vasilescu \cite{fang2024novelty}. 
Hence, we adopted a definition of novelty, which entail elements of uniqueness and originality \cite{lloydcox2022evaluatingcreativity}, see Appendix  \ref{app:prompt} for the full prompt:
\begin{quote}
\textit{How unique and original is the project's concept, approach, or solution? Does it introduce new ideas, methods, or perspectives that are significantly different from existing ones?}
\end{quote}
To operationalize this definition for the analysis, we decided to replicate the approach by Fang, Herbsleb and Vasilescu who defined this ``as a function of the libraries and packages a project imports [i.e.] projects built on top of more atypical combinations of libraries are considered to be more innovative'' \cite{fang2024novelty}.
We replicated this approach and identified the hackathon projects which included unusual combinations of packages and libraries. 
We included the five most popular programming languages used in hackathon projects in our dataset: Python, Node.js, Java, C\#, and PHP.
While this may, indeed, exclude some potentially creative projects, we decided for this approach partly to repeat the research by Fang, Herbsleb and Vasilescu \cite{fang2024novelty} but also because we hypothesized that this approach includes projects that have a higher chance of getting utilized or continued in the future because the languages they were developed in are so wider spread.%versatile projects as these languages are used for many different types of applications.
%While this will exclude some potentially creative projects which do not use Python or even include code at all, we decided for this approach partly to repeat the research by Fang, Herbsleb and Vasilescu and because of Python's popularity:
%Python has been one of the fastest-growing programming languages for years \cite{srinath2017python}, and is one of the most used programming languages \cite{statistaInfographicPython}, also in hackathons in particular \cite{techcrunchWhichProgramming}.
%Since Python is a very versatile programming language, by targeting it we hypothesize that we will cover as versatile hackathon projects built on Python. 

\paragraph{Usefulness}
Luchini and colleagues do not use the term ``usefulness'' in their prompt design \cite{luchini2023automatic}, however, as there is a lack of established terminology in this area, concepts used in definitions of creativity may therefore vary in their exact wording.
Hence, Luchini and colleagues include the terms ``completeness'' and ``effectiveness'', which closely resembles usefulness: ``Completeness refers to how well the solution addresses multiple issues raised by the problem. Effectiveness examines whether the solution is viable, feasible, practical, or appropriate'' \cite{luchini2023automatic}.
Our adopted definition of usefulness is the following (see also Appendix  \ref{app:prompt}):
\begin{quote}
\textit{How practical and appropriate is the project in addressing the problem, situation, or challenge its targets? Does it effectively solve a real-world issue or meet a specific need?}
\end{quote}
We first needed to distinguish a subset of the dataset which can be identified as containing creative projects.
Since it would not be feasible to conduct expert evaluation on the full dataset, we opted for a creativity proxy, by considering winning hackathon projects, which include a \textit{winner}-tag in our dataset.
This, of course, excludes hackathon projects which did not win a certain hackathon according to a set of criteria, but for our exploration of how to evaluate creativity at scale we find this a sufficient proxy for the expert evaluation.
While we do not know at scale which criteria define winner-projects, considering the wide-spread emphasis on creativity in hackathons as mentioned in the introduction, we hypothesize that the majority of our dataset's winner-projects have been considered creative or have addressed a challenge in a sufficient way which has been perceived as useful. 

However, what is considered useful within a hackathon context may not necessarily be considered useful outside of the hackathon context. 
We therefore hypothesized that how \emph{accessible} a hackathon project is for outsiders can be a proxy for considering usefulness outside of a hackathon context. 
Operationalizing this hypothesis, we decided to follow an approach inspired by Imam and colleagues who explored antecedents of reuse of code that was created during hackathons \cite{imam2021secretlifeofhackathoncode}.
If code is reused this points towards the code being useful outside of the confines of a hackathon. Imam et al. found that the availability of data and documentation in addition to code and the presence of an open-source license significantly contributed to a project being reused. 
Furthermore, larger data files may indicate that projects are structured with reusability in mind.
Following their findings, we include hackathon projects which have GitHub repositories and include data files or folders and an Open Source license for the repository in our final subset of creative perceptible hackathon projects.

%GitHub accessibility, licensing, code scripts, and non-code data are essential factors influencing the future usability and reference value of hackathon projects. 
%Accessible GitHub links allow projects to be reviewed and utilized over time, ensuring their long-term relevance. 
%, enhancing their utility and ease of use for others. 
%As a result, both accessibility and the scale of data play vital roles in maintaining the relevance and usefulness of hackathon projects.

Hence, the subset of creative hackathon projects satisfies the above-mentioned criteria of novelty (including atypical combinations of software) \textit{and} usefulness (containing a winner-tag \textit{and} is accessible for outsiders). 
We discuss the limitations of this delimitation in the Discussion section.
Using our operationalized definitions of novelty and usefulness, we identified \textbf{619} \textbf{creative perceptible products in the hackathon context}, see fig. \ref{fig:filtering}.
The technical details of this process are described in section \ref{sec:setup}.
The next subsection describes some guiding questions for the different angles which we analyzed the dataset from.

\paragraph{The Interaction Between Aptitude, Process and Environment:}
We finally explored the three \textbf{aspects of creativity} in terms of aptitude, process and environment in this subset of the creative perceptible hackathon products.
However, we found that the dataset contained rich data which went beyond these three aspects, therefore we extended the framework to better suit our analysis.
Some frameworks have been suggested to distinguish between different aspects of creativity such as: The 4P model \cite{rhodes1961AnAnalysisofCreativity} (People, Process, Product, Press), the 5A model \cite{glavenau2013FiveAsFramework} (Actor, Action, Artifact, Audience, Affordances) and the 7C model (Creators, Creating, Collaborations, Contexts, Creations, Consumption, Curricula) \cite{lubart2017The7CsofCreativity}. 
We turned to the framework of the 7 C's of creativity because it has the added aspect of Collaboration, which is an important characteristic part of hackathons \cite{falk2024future}.
We excluded Creating and Curricula from our analysis for the following reasons:
Creating would require more in-situ data which documents the creative design process in terms of, for example, the moment to moment decisions. 
The dataset does not contain this kind of data.
Similarly, Curricula revolves around the overarching context of how creativity is taught and learned in the specific context. 
This leaves us with the aspects of \textbf{Creators, Collaborations, Contexts, Creations}, and \textbf{Consumption}.

With this framework, we were able to approach the computational analysis of creativity in hackathons in a structured way and base the analysis on established definitions from creativity research.
Turning back to the standard definition of creativity, the findings in section \ref{findings} then relate to the \textbf{interaction} between the selected aspects of the 7 C model \cite{lubart2017The7CsofCreativity}.
In the following subsections, we outline research questions for each of the five selected C's.

\textbf{Creators}:
Lubart describes creators as referring to ``those who engage in the production of original, meaningful content'' \cite{lubart2017The7CsofCreativity}.
To explore the dataset, we asked the following questions which could provide insights about the creators of the creative hackathon projects: 
\begin{enumerate}
    \item Is the creator a hackathon winner?
    \item Which skill sets have they provided on their profile?
    \item What is the experience of creators? How many hackathons have they participated in? How many years have they participated in hackathons?
\end{enumerate}

\textbf{Collaborations} ``is the term used to signify the involvement of significant others in the creative process. This may be an individual creator, such as a writer, who interacts with another person, such as his or her literary agent or critic, it may be a dyad of creators who work together, or a team of people who work on a project [...] The collaboration, in terms of interaction patterns, the nature and complementarity of the collaborators (team diversity) are some specific examples of topics that concern this `C'.'' \cite{lubart2017The7CsofCreativity}
From the dataset, we wanted to explore:
\begin{enumerate}
    \item Do creators work together in different hackathons?
    \item What are typical team sizes?
    \item What skills and interests do teams possess?
\end{enumerate}

\textbf{Contexts} ``refers to the physical and social world in which creators engage in the creative process [...] The environment provides resources and constraints, it orients behavior. It affords certain actions more than others, facilitating or hindering creative behavior but also providing the field within which new productions will be situated and evaluated'' \cite{lubart2017The7CsofCreativity}.
To explore the hackathon contexts, we looked for the following elements:
\begin{enumerate}
    \item Which hackathon themes do creative projects address?
    \item What is the distribution of themes?
    \item What is the level of competition in hackathons, i.e., what is the interaction between the number of projects of hackathon and the number of winners in a hackathon (is there only one first prize or multiple prizes)?
    \item How big are hackathons with creative projects?
    \item Are the hackathons mostly onsite or online?
    \item How many creators participate?
\end{enumerate}

\textbf{Creations} are ``the production resulting from the creative process, maybe a tangible or intangible output. It may be a relatively unformed idea, or a full-fledged `product'. The characteristics of the production, such as its originality compared to previous works, and its `usefulness', maybe some criteria that the creator and external judges take into account'' \cite{lubart2017The7CsofCreativity}.
To explore the creations, we wanted to analyze the following components:
\begin{enumerate}
    \item How does the hackathon projects compare their inspiration source with what they do? What is the \textit{semantic distance} within project descriptions, i.e. ``an aspect of originality [or novelty] reflecting remote relationships between concepts'' \cite{luchini2023automatic}.
    \item  How elaborate are the project descriptions?
    \item How much do project descriptions align with or differ from the hackathon context, i.e. the posed challenge the project was meant to address?
\end{enumerate}

\textbf{Consumption} ``refers to the adoption of creative ideas and productions. Those who encounter a creative product may adopt it more or less quickly, with more or less enthusiasm. Creations are situated within a context of the marketplace of existing ideas, products, or previously known solutions'' \cite{lubart2017The7CsofCreativity}.
As we described above, a hackathon project may be perceived as creative in one way during the hackathon and in another outside the hackathon context. 
To explore whether creative hackathon projects are adopted outside the hackathon context, we explore the following: 
\begin{enumerate}
    \item Do the projects have GitHub repositories, and are these reachable and maintained?
    \item What is the life span of a project outside of the hackathon?
\end{enumerate}

While the findings in section \ref{findings} do not reflect all these research questions for the C's, they were used to structure the initial analyses of the large-scale dataset and the results of these initial explorations can be found in the appendix, see section \ref{appendix:correlation}.
Researchers who are interested in similar evaluation methods for creativity, may find the tables interesting to see which parameters may not contribute to creative hackathon projects.

\subsection{Experimental Setup\label{sec:setup}}
\begin{figure}[t]
    \centering
    \includegraphics[width=\linewidth]{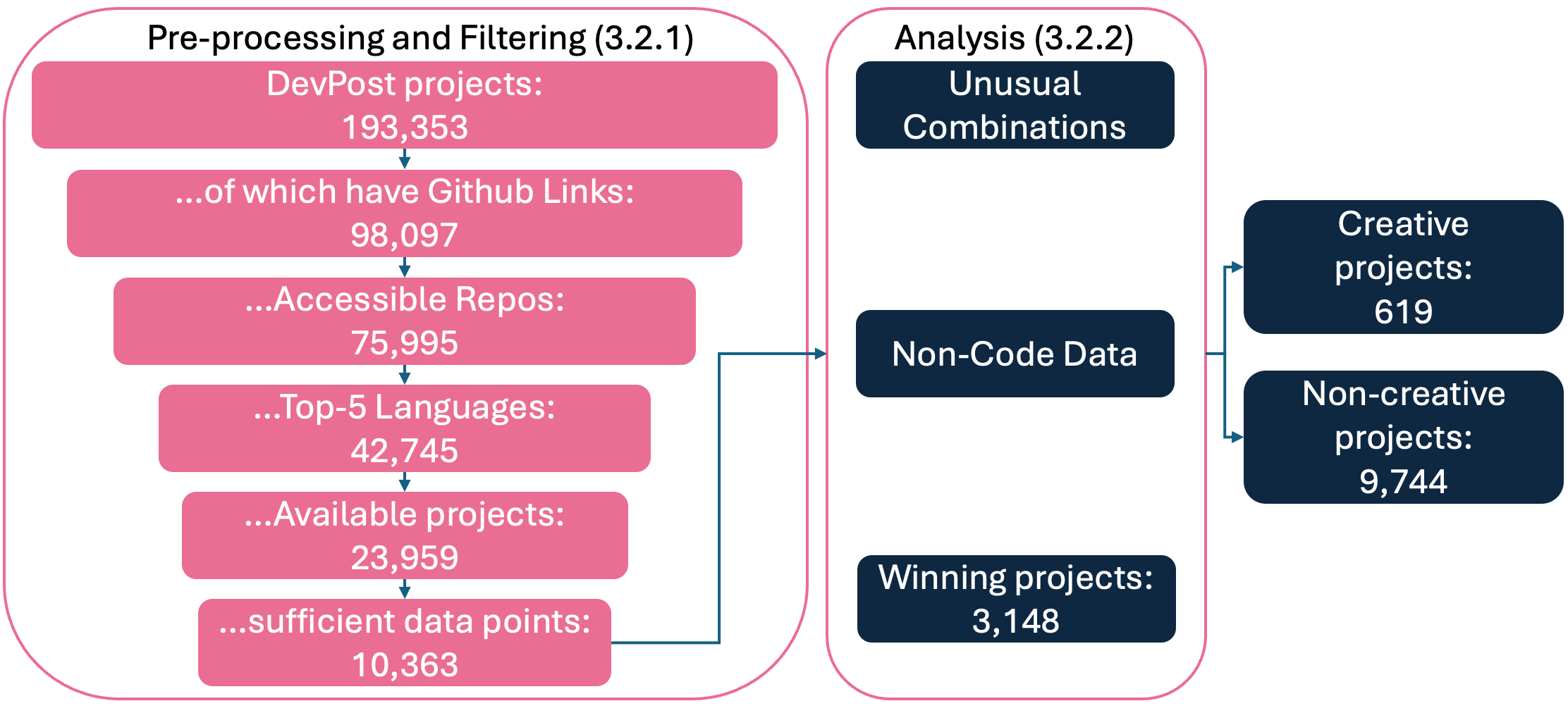}
    \caption{Number of projects remaining after each pre-processing step (left). This process is described in detail in section \ref{Data collection and pre-processing}; Extracted variables for analysis (mid). The analysis is described in detail in section \ref{novelty}; Resulting prediction of creative projects (right). The analysis results are described in detail in section \ref{predicting creative projects}.}
    \label{fig:filtering}
    \Description{This figure illustrates a data processing and analysis workflow divided into two main sections: Pre-processing and Filtering (3.2.1) and Analysis (3.2.2). The pre-processing flow starts with 193,353 DevPost projects, of which 98,097 have GitHub links. From these, 75,995 repositories were accessible, and 42,745 used the top-5 programming languages. The number of available projects was further reduced to 23,959, and finally, 10,363 projects had sufficient data points. In the Analysis section, these projects were categorized into "Unusual Combinations", "Non-Code Data" categories and 3,148 projects were identified as winning projects. We use the criteria of novelty and usefulness to further split these projects into creative projects (619) -- which fulfilled the categories ``Unusual Combinations'', ``Non-Code Data'' and were winning projects -- and non-creative projects (9,744) -- which did not fulfill these three categories. The diagram uses pink boxes for the pre-processing steps and dark blue boxes for the analysis outcomes, with arrows indicating the flow of data processing.}
\end{figure}

In this section, we first describe the series of steps in our pipeline for data collection and pre-processing, before analyzing this data with regards to our operationalizations of usefulness and novelty, and finally predict the creativity of projects with a mixed random effects model.
As each stage in the data pre-processing process, relating to the five C's, requires adding incremental filtering criteria, the number of projects that we analyze varies as described in Fig.~\ref{fig:filtering}.

\subsubsection{Data Collection and Pre-processing}
\label{Data collection and pre-processing}
We collected publicly available information on hackathons, projects and participants from the hackathon database Devpost which represents a snapshot of the state of hackathons, projects and participants at the moment in time when we collected the data. Historical data was not available to us.
We begin by curating and processing a dataset with 193,353 projects (ca.~6k hackathons and 314k participants). 
After preprocessing, we narrowed it down to 23,959 projects with GitHub repositories. 
Further filtering for repositories with code scripts importing more than two packages and non-code data yielded 10,363 projects for usefulness and novelty analysis, in which 3,148 projects are identified as winning projects.
Finally, the projects are identified into 619 identified as creative -- which fulfills the categories ``Unusual Combinations'', ``Non-Code Data'' and were winning projects --  and 9,744 as non-creative (Fig.~\ref{fig:filtering}).

\begin{figure}[t]
    \centering
    \includegraphics[width=\linewidth]{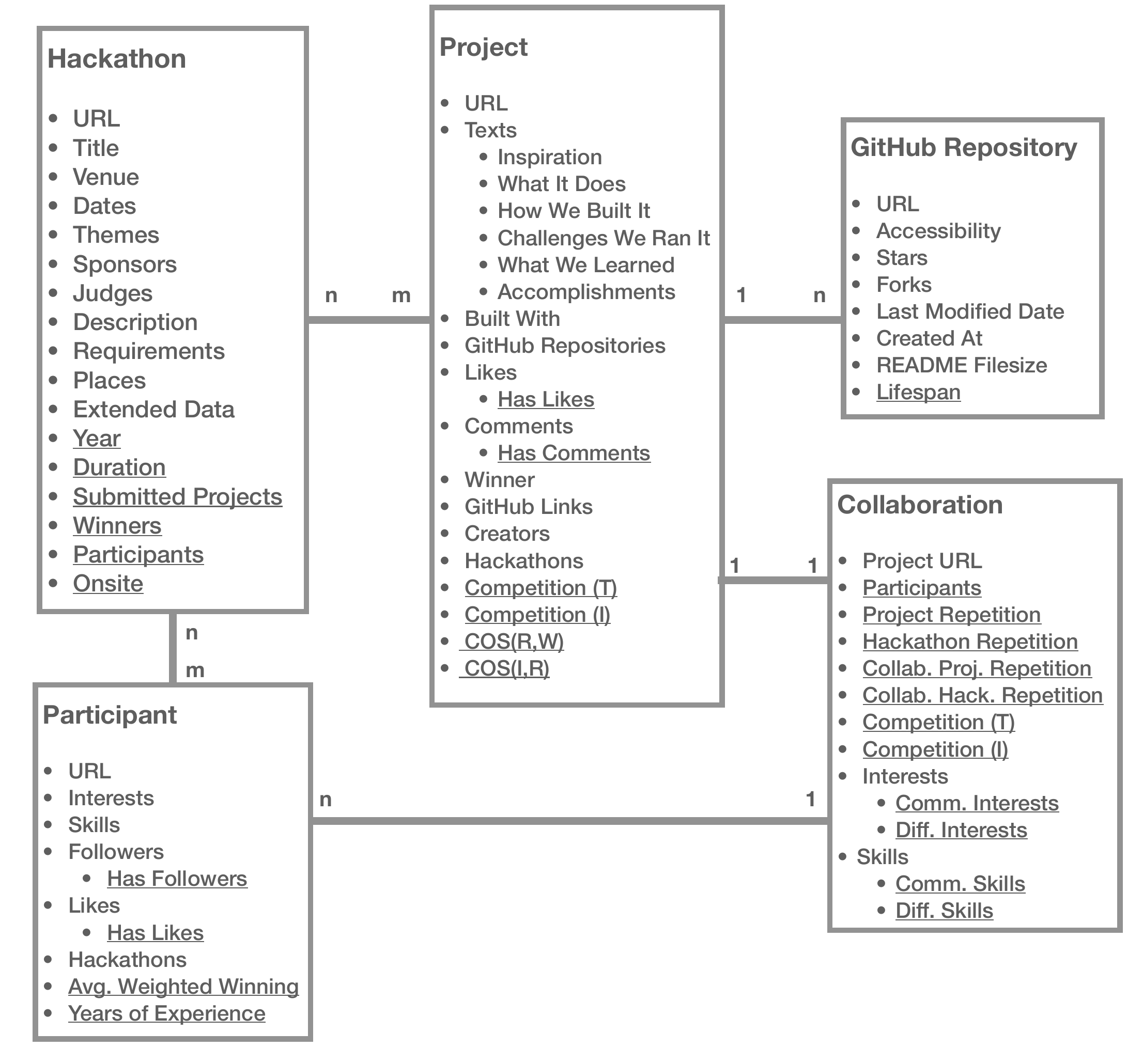}
    \Description{This figure illustrates the variables in the analyzed dataset, encompassing hackathons, projects, participants, and GitHub repositories associated with the projects, along with the newly constructed collaboration dataset, which captures interaction metrics of participants at the project level. It also depicts the relationships between each dataset: for example, a hackathon can be related to multiple projects, a project can have multiple GitHub links, a project has a one-to-one relationship with its collaboration metrics, hackathons relate to participants in an n-to-m manner, and participants within a project have a one-to-many relationship with collaborations. }
    \caption{Variables in our dataset. The underlined \underline{Variables} are processed from the collected data. }
    \label{fig:dataset_variables}
\end{figure}

As shown in Fig.~\ref{fig:dataset_variables}, each project can be submitted to multiple hackathons, and conversely, a single hackathon can host multiple projects.
To investigate collaboration patterns in hackathons, we process the dataset to capture the dynamics of collaboration among creators at the project level, considering both their experience in hackathon participation and a variety of individual interests and skill sets.
In the following, we explain the pre-processing steps in detail for exploring each aspect of the five C's.

% \subsubsection{Novelty}
% We begin by exploring the aspect of \textit{novelty} in our sample of hackathon projects, following Fang, Herbsleb and Vasilescu \cite{fang2024novelty}. 
% For this, we curated  
% %23,959 
% GitHub repositories associated with the projects built with the five most popular programming languages in our sample: Python, Node.js, Java, C\#, and PHP. 
% From these, we extract code, documentation, data and metadata such as stars, forks, and timestamps and measure the atypicality of package combinations using Monte Carlo simulation (cf. Section~\ref{novelty}).
% %In comparison to Fang, Herbsleb and Vasilescu's dataset,  our approach results in a smaller dataset.

%Among the projects, 140,608 include text descriptions, which are further used to calculate semantic distance to explore the association of projects and hackathons. 

\paragraph{Creators}
%As a number of webpages for hackathon participants are no longer available, w
We crawled data for $310,900$ creators which have participated in the hackathon projects, observing variables including interests, skills, received likes, had followers and participated in 1.5 hackathons and projects on average. %how many hackathons they participated in. 
%However, data such as non-zero number of followers and likes are sparse among participants, i.e., 70,363 and 48,584, respectively. 
As very few participants have followers or likes, we binarize these numerical data points into \textit{have followers} and \textit{have likes} for further analysis.
To explore the correlation between creators' individual traits and their winning hackathons, a new variable is calculated as follows:
\begin{equation}
    \text{AVG. Weighted Winning}= \frac{1}{n}\sum^{n}_{i=1} \frac{\text{Places}_i}{\text{Participants}_{j} * \text{Sub. Projects}_{i}},
\end{equation}
where $n$ is the number of hackathons that the creator won, 
Places$_i$ is the number of places reserved for winners in a hackathon $i$, and 
Participants$_j$ is the number of participants who are creators in one project team $j$, and Sub. Projects$_{i}$ is the total number of submitted projects in the regarding hackathon $i$.
Table~\ref{tab:corr_creators} contains the correlations among variables for participants.

\paragraph{Collaborations}
Collaboration among creators is a dynamic temporal process.
The collaboration among creators in hackathons accumulates through the repeated participation of the same pair of participants, so does the collaboration on the projects. 
Moreover, individual experience accumulates through repeated participation in projects and hackathons with others.
To capture the dynamics of collaboration for each year, we devise an algorithm~\ref{algo:collab_interaction}. 
It initializes dictionaries to track creator-to-creator interactions for projects and hackathons, as well as individual hackathon and project repetition.
Then, it iterates through projects by year, updating the interaction counts for each pair of creators who collaborate on a project or participate in the same hackathon. 
Finally, it computes the interaction metrics for each creator at both the project and the hackathon level.
The results are then aggregated over time, recorded as \textit{Collaboration Project Repetition}, \textit{Collaboration Hackathon Repetition}, \textit{Project Repetition} and \textit{Hackathon Repetition}.
Furthermore, we extract the intersected, different, and all sets of \textit{interests} and \textit{skills} among participants for each project as variables to explore the dynamics of collaboration.
As shown in Table~\ref{tab:corr_collab}, the variety of skills is positively correlated with repeated collaborations for projects and hackathons. 

% findings: 

% \begin{itemize}
% \item interaction and experience correlate positively with the team size, as it should be.
%     \item common interests do not compell winner projects.
%     \item project /hackathon interaction and experience are positively correlated, 
%     \item in general, more collaboration and variety of interests is conducive to winning, but not that strongly correlated.
%     \item winner projects correlate with higher number of overall skillsets and also different skillsets
    
% \end{itemize}

\paragraph{Contexts}
A total of 5,458 hackathon URLs remain accessible, and information has been extracted from them for the period 2009-2024 in the dataset. 
Using the crawled data, we extract the \textit{year} and \textit{duration} by days of the hackathons by parsing the schedule dates using Python package \textit{dateutil}.\footnote{\url{https://pypi.org/project/python-dateutil/}} The number of submitted projects, the number of actual winners and participants for each hackathon are aggregated utilizing project data. 
Hackathons occur worldwide, and we are particularly interested in whether the format of the hackathons---whether online or onsite---affects creativity. To analyze this, we convert the \textit{venue} variable into a binary variable, \textit{onsite}.
Furthermore, to measure the level of competition and explore its relevance in the context of hackathons, we calculate the ratio of reserved winning spots to the number of participating teams, as well as to the number of participants. This provides \textit{competition} metrics at both the team level (T) and the individual level (I). Refer to Table~\ref{tab:corr_contexts} in Appendix~\ref{appendix:correlation} for correlations among variables in hackathons.

\paragraph{Creations}
Projects are creations in the context of hackathons. There are originally 193,353 project data in total, with some projects submitted to multiple hackathons. And 187,425 of which contain information descriptions.
To explore how hackathon project descriptions, their inspiration and the overarching hackathon description compare, we calculate the semantic distances by leveraging the text descriptions from fields such as \textit{What It Does}, \textit{Requirements}, and \textit{Inspiration}. %
To achieve this, we use a sentence transformer \textsc{LaBSE}\footnote{huggingface: sentence-transformers/LaBSE} to encode texts into 768-dimensional representation vectors, with maximal sequence length 256. %, and the resulting sentence embeddings are in dimension 768. 
We then calculate cosine similarities between (1) the embeddings of \textit{Inspiration} and \textit{What It Does} and (2) the embeddings of \textit{What It Does} and \textit{Requirements}, recorded as $COS_{R, W}$ and $COS_{I,W}$, respectively.

% x
% - [ ] to compare “inspiration” with “what it does” (within project) $COS_{I,W}$
%     - do they use a lot of space for describing inspiration?
%     - token counts.
% - semantic distance 
% - [ ] “what it does” and “requirements”(hackathon)
%         - perceived distance???
%     - [ ] “inspiration” and “what it does” ( how much group level obeys the hackathon)
%         - correlate with creativity/innovation scores???
% - popularity
%     - [x] likes and comments from DevPost.

 % perhaps go to appendix.
 % \begin{figure}[!h]
 %     \centering
 %    \includegraphics[width=0.4\linewidth]{images/cos_what_insp.pdf}
 %     \includegraphics[width=0.4\linewidth]{images/cos_what_req.pdf}
 %     \caption{Distributions of Semantic Distances of }
 %      \Description{TODO}
 %     \label{fig:enter-label}
 % \end{figure}

\paragraph{Consumption}
As shown in Fig.~\ref{fig:filtering}, among the 193,353 projects we are investigating, 98,097 of them have GitHub links, among which 75,995 GitHub repositories are still accessible, 42,745 of which are written in the top 5 programming languages, and eventually, we were able to crawl 23,959 GitHub repositories with non-empty scripts,  4,037 of which has a specific license, the top three of which are MIT, Apache and GNU licenses (Fig.~\ref{fig:license}). Moreover, 22,718 GitHub repositories contain non-code data (Fig.~\ref{fig:datasize_github_distribution}).
As consumption overlaps with the number of retrievable repositories, this coincides with the number of projects we use for the remainder of our analysis.

%%%TODOL

The Pearson correlation coefficient between the winning project and GitHub accessibility is 0.06 with $p<0.001$.
%GitHub accessibility, licensing, code scripts, and non-code data are essential factors influencing the future usability and reference value of hackathon projects. Accessible GitHub links allow projects to be reviewed and utilized over time, ensuring their long-term relevance. Furthermore, larger data files may indicate that projects are structured with reusability in mind, enhancing their utility and ease of use for others. As a result, both accessibility and the scale of data play vital roles in maintaining the relevance and usefulness of hackathon projects.

% \begin{figure}[!h]
%     \centering
%     \includegraphics[width=0.6\linewidth]{images/github_assessibility.pdf}
%     \caption{GitHub Accessibility by Year of the Hackathon.}
%     \label{fig:github_assessibility}
%     \Description{}
% \end{figure}

\subsubsection{Analysis of Usefulness and Novelty}\label{novelty}

To assess both usefulness and novelty, we adopt a two-pronged approach. 
Usefulness is evaluated by focusing on hackathon projects that include non-code data and those recognized as winners, reflecting their practical application and broader impact beyond hackathons, extending Fang, Herbsleb and Vasilescu's work \cite{fang2024novelty}.
Novelty, on the other hand, is measured by unusual combinations of imported packages within the project code, highlighting innovative approaches to problem-solving.

\textbf{Usefulness: Non-code data and Winners}
% To extend Fang, Herbsleb and Vasilescu's work, we add our operationalization of usefulness to the analysis. 
Continuing the dataset of 23,959 Github repositories in total, with Python, Java, Node.js, C\# and PHP in the \texttt{built with} tag from the project. 
In detail, we extract information such as \textit{data size (in Bytes)}, \textit{License}, \textit{GitHub Lifespan (Days)}, \textit{\# Data Files}, \textit{\# Data Folders}, \textit{Watches} and \textit{Stars}.

\textbf{Novelty: Unusual combinations}
In addition, to operationalize novelty of the project, we investigate the unusual combination of the imported packages in the code.
To achieve that, we process the same set of GitHub repositories with the following steps: (1) the utilized packages are extracted from each script;
% excluding the standard libraries, and local packages;
% \footnote{Using Python package~\url{https://github.com/Nicolas-Reyland/findpydeps}}; 
(2) compute the number of imported packages for the same project;
(3) filter the repositories out when their imported packages are fewer than 2.
As a result, we have 10,363 GitHub repositories for analyzing novelty leveraging the atypical combinations of imported Python packages.

%We adopt the method by Fang, Herbsleb and Vasilescu who quantified novelty as project atypicality from~\cite{fang2024novelty}. 
%In comparison,  we have much smaller dataset compared to their work.

To conduct our experiment, first, we simulate the actual package imports over time. We represent package imports by projects over time as a matrix where each row is a project, and each column is a package. In this matrix, a value of $1$ indicates that a project imports a particular package in that year, while $0$ signifies no import.
Then for each year, we simulate random package combinations for projects. Each project imports the same number of packages as in the actual data, but the choice of which packages are imported is random. This gives us a counterfactual dataset to compare against. This uses the Monte Carlo simulation to capture the atypicality of package combinations.
Then \textit{empirical frequency} is computed, where how often two specific packages are used together in the actual data.
At the same time, we compute \textit{simulated frequency}, which calculates how often the two packages appear together by chance after several simulation runs.
The atypicality score is derived from the ratio of empirical frequency to the average simulated frequency. The combination is atypical if the empirical frequency is much lower than expected.
We compute a z-score for each package with the equation:
\begin{equation}
    z_{ijt} = (obs_{ijt} - exp_{ijt})/\sigma_{ijt}
\end{equation}

where $obs_{ijt}$ represents the empirically observed frequency of packages $i$ and $j$ appearing in the same project in year t, $exp_{ijt}$ is the average number of times  that packages $i$ and $j$ appear in the same project in the year $j$ over twenty simulated event sets, and $\sigma_{ijt}$ is the standard deviation of the co-occurrence frequency of packages $i$ and $j$ in those sets as well.
In the end, we use the smoothed version of z-score to measure atypicality, as follows:
\begin{equation}
Z_{ijt}=
\begin{cases} 
\log(z_{ijt} + 1) & \text{if } z_{ijt} \geq 0 \\
- \log(-z_{ijt} + 1) & \text{if } z_{ijt} < 0
\end{cases}
\end{equation}
A low Z-score indicates high atypicality and, consequently, novelty, whereas a high Z-score suggests the opposite.

% z-score: atypicality score, the higher, the less novelty.

% The correlation indicates that there is a weak correlation between novelty and winner projects. 

%4691 projects, with python code crawled, of which 1,286 are winning projects. 

\subsubsection{Predicting Creative Projects using Mixed Random Effects Logistic Regression}
\label{predicting creative projects}
To investigate novelty and usefulness as proxies for creativity of projects, we use a combination of Z-score and \textit{winner}-tag as the dependent variable in a mixed-effects logistic regression model. 
This approach allows us to identify which variable contribute most to creative projects. 
% conducting mixed effects logistic regression modeling to explore which variables are more conducive to creative projects.
The Z-score is a continuous variable, where a Z-score of zero indicates a high level of novelty, while \textit{winner}-tag is binary. 
We integrate those two variables into a binary variable using the following equation:
\begin{equation}
\text{Creative-Projects} =
\begin{cases}
1, & \text{if \textit{winner}-tag} =1 \ \text{and} \ Z\text{-score} = 0 \\
0, & \text{otherwise}
\end{cases}
\end{equation}

To simulate the conditions in which a project qualifies as a creative project, characterized by the presence of a \textit{winner} tag and a valid Z-score, we filter the data to include only hackathons that have at least one designated \textit{winner}-tag. 
A project may have different outcomes across various hackathons (either winner or not), but it retains the same Z-score for its GitHub repository. 
In total, we have a dataset of 10,363 projects for further analysis, with 619 identified as creative and 9,744 projects identified as ``un-creative''.
In the following we analyze what distinguishes the 619 creative projects from the rest.

We process the variables from various aspects of the dataset (containing both creative and un-creative projects) - namely participants, projects, and hackathons - into project-level variables.
For example, the number of participated hackathons and the years of experience for participants are averaged for the regarding project. 
To ensure the comparability among variables, numerical variables are scaled. 
We then apply a mixed effects logistic regression model using \textit{glmer} function from R-package \textit{lme4}.
In this model, variables directly related to hackathons are treated as random effects, where other variables are fixed effects. The model uses \textit{Binomial} family and is optimized with \textit{nloptwarp} control optimizer. 
We examine the multi-collinearity of the variables in the models, and select the variables which are the most relevant and non-redundant. The results are presented in Table~\ref{tab:regression}, and the heatmap of Spearman's rank correlation coefficients among the selected variables are reported in Fig.~\ref{fig:correlation_mixed_effects}.

\begin{table}[t]
    \centering
    \caption{Model predicting Creative Projects. p<0.1 (.), p<0.05 (*), p<0.01 (**), p<0.001 (***).
}
    \resizebox{0.8\linewidth}{!}{
    \begin{tabular}{l|cccc}    
   % (Intercept) &  -3.01431 (***)  &  0.06452 & -46.722  & < 2e-16 \\
  \toprule
        (Intercept)  &  -3.250649 (***)  &  0.137272& -23.680 & < 2e-16 \\
       \midrule
      % \underline{ Creators} & \\
    \textbf{Hackathon}  & & & & \\
    % \textit{  \#Participants (in Hackathons)} & -0.22399 (*) & 0.11262 & -1.989 & 0.04671  \\
   \textit{  \#Participants (in Hackathons)}  & -0.226608 (*)  & 0.113896  & -1.990  & 0.04664 \\
   \textit{  Onsite} &  0.067976  &  0.065501 &   1.038 & 0.29937 \\
   \textit{  Duration}  & -0.078477 &   0.057713 &  -1.360 & 0.17390    \\
   \midrule
    \textbf{Team} & & & & \\
    \textit{  Competition} &  0.424168 (***) & 0.102305  &  4.146  & 3.38e-05 \\
    \textit{  \# Team Members}  &  0.156271 (*) &  0.073603   &  2.123  &  0.03374   \\
%    \textit{  \#Participants in (Project)} & 0.14238 (.) & 0.07398 & 1.925 & 0.05428 \\
    \textit{  Different Interests} &  -0.305535 (*) &   0.131234  &  -2.328 &  0.01990 \\ 
    \textit{  All Skills} &     0.168547  & 0.154694  & 1.090  & 0.27591  \\
    \textit{  All Interests} &   0.121256  &   0.128110   &   0.946   &  0.34389  \\
    \textit{  Hackathon Repetition} &    -0.046386 &   0.097278 &   -0.477  &  0.63347 \\
    \textit{  Different Skills} &  -0.213552 &  0.173468 &  -1.231  & 0.21829  \\
    \midrule
    \textbf{Project}  & & & & \\
    \textit{  Has Likes} &  0.275278 (.)  &  0.152604  & 1.804 & 0.07125  \\
    \textit{  GitHub Watches} & 1.584174 (*) & 0.654169  & 2.422 &  0.01545   \\
    \textit{  GitHub Data Size} &  0.097405 (**) &  0.031475  &  3.095  &  0.00197  \\
     \textit{Project Repetition} &   0.021523  &  0.120655 &    0.178 &   0.85842    \\
             \textit{  COS(I,W)} & 0.008229 &  0.056720  & 0.145 & 0.88464 \\
    \textit{  Has Comments} &  0.023240  & 0.150865  & 0.154 & 0.87757   \\
    \textit{  GitHub has License} & 0.050863  & 0.043701  &  1.164  & 0.24447    \\
    \textit{  \# GitHub Data Folders} & -0.009275   & 0.123769  & -0.075  & 0.94027    \\
    \textit{  GitHub Lifespan (Days)} &  -0.077932  &  0.048979 &   -1.591 &   0.11158 \\
        \textit{  COS(R,W)} &  -0.095202    & 0.060317 &  -1.578 &  0.11448    \\
    \textit{  \# GitHub Files} &  -0.643382 &  0.398770  & -1.613 & 0.10665  \\
    \textit{  GitHub Stars} &  -11.130124 &   7.239957 &  -1.537 &  0.12422    \\
    \midrule
    \textbf{Individual} & & & & \\
    \textit{  Average Won Hackathons} &  0.918309 (***) &  0.158897  &  5.779 & 7.50e-09 \\
    \textit{  Hackathon Repetition} &  0.484675 (***) & 0.058513 &  8.283 &  $<$ 2e-16 \\
    \textit{  Has Followers} & 0.246149 (***) &  0.055379 &   4.445 & 8.80e-06 \\
    \textit{  Average number of Hackathons} & -0.769460 (***)  &   0.192835 &   -3.990 &  6.60e-05 \\
    \textit{  Project Repetition} &  -0.247321  (*) &  0.100642 & -2.457 & 0.01399  \\
    \textit{  Competition} &   -0.220472 (*)  & 0.105092 & -2.098  &0.03591 \\

    \textit{  Has Likes} &  -0.059746 &   0.056079  &-1.065  & 0.28670    \\
        
    \midrule
     & \textbf{$R^2_{m}$ } &   \textbf{0.967} &  \textbf{$R^2_{c}$ }  & \textbf{ 0.970} \\
                \bottomrule
        
    \end{tabular}}
    \Description{This statistical table presents results from a mixed-effects regression model predicting Creative Projects, demonstrating exceptional explanatory power with marginal (R²_m = 0.967) and conditional (R²_c = 0.970) R-squared values. Beginning with a significant negative intercept of -3.250649 (p < 0.001), the analysis spans four categories: Hackathon, Team, Project, and Individual characteristics. In the Hackathon category, the number of participants shows a significant negative effect (-0.226608, p < 0.05), while onsite participation (0.067976, p = 0.29937) and duration (-0.078477, p = 0.17390) remain non-significant. Team characteristics reveal that Competition has a strong positive effect (0.424168, p < 0.001), Has Likes shows a marginally positive effect (0.275278, p < 0.1), and Number of Team Members demonstrates a positive influence (0.156271, p < 0.05), while Different Interests shows a significant negative effect (-0.305535, p < 0.05). Project metrics indicate that GitHub Watches (1.584174, p < 0.05) and GitHub Data Size (0.097405, p < 0.01) have significant positive effects, while other GitHub metrics, including Stars (-11.130124, p = 0.12422), Files (-0.643382, p = 0.10665), and Lifespan (-0.077932, p = 0.11158), remain non-significant. Individual factors demonstrate strong effects, with Average Won Hackathons (0.918309, p < 0.001), Hackathon Repetition (0.484675, p < 0.001), and Has Followers (0.246149, p < 0.001) showing significant positive influences, while Average number of Hackathons (-0.769460, p < 0.001), Project Repetition (-0.247321, p < 0.05), and Competition (-0.220472, p < 0.05) demonstrate significant negative effects. The comprehensive model suggests that while individual experience and team competition positively influence creative projects, factors such as over-participation in hackathons and diverse interests within teams may have detrimental effects.}
    \label{tab:regression}
\end{table}

\begin{figure}
    \centering
    \includegraphics[width=\linewidth]{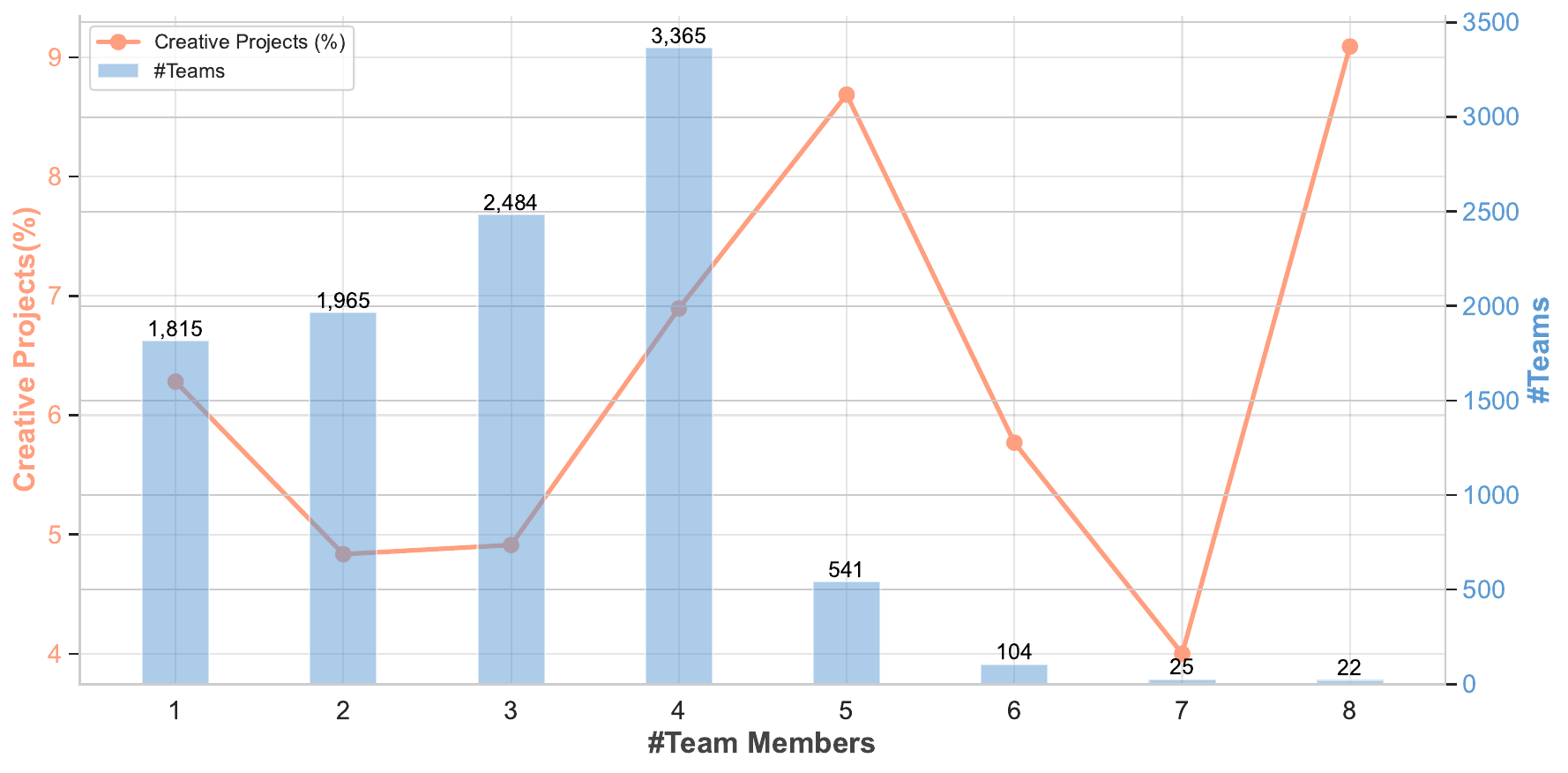}
    \includegraphics[width=\linewidth]{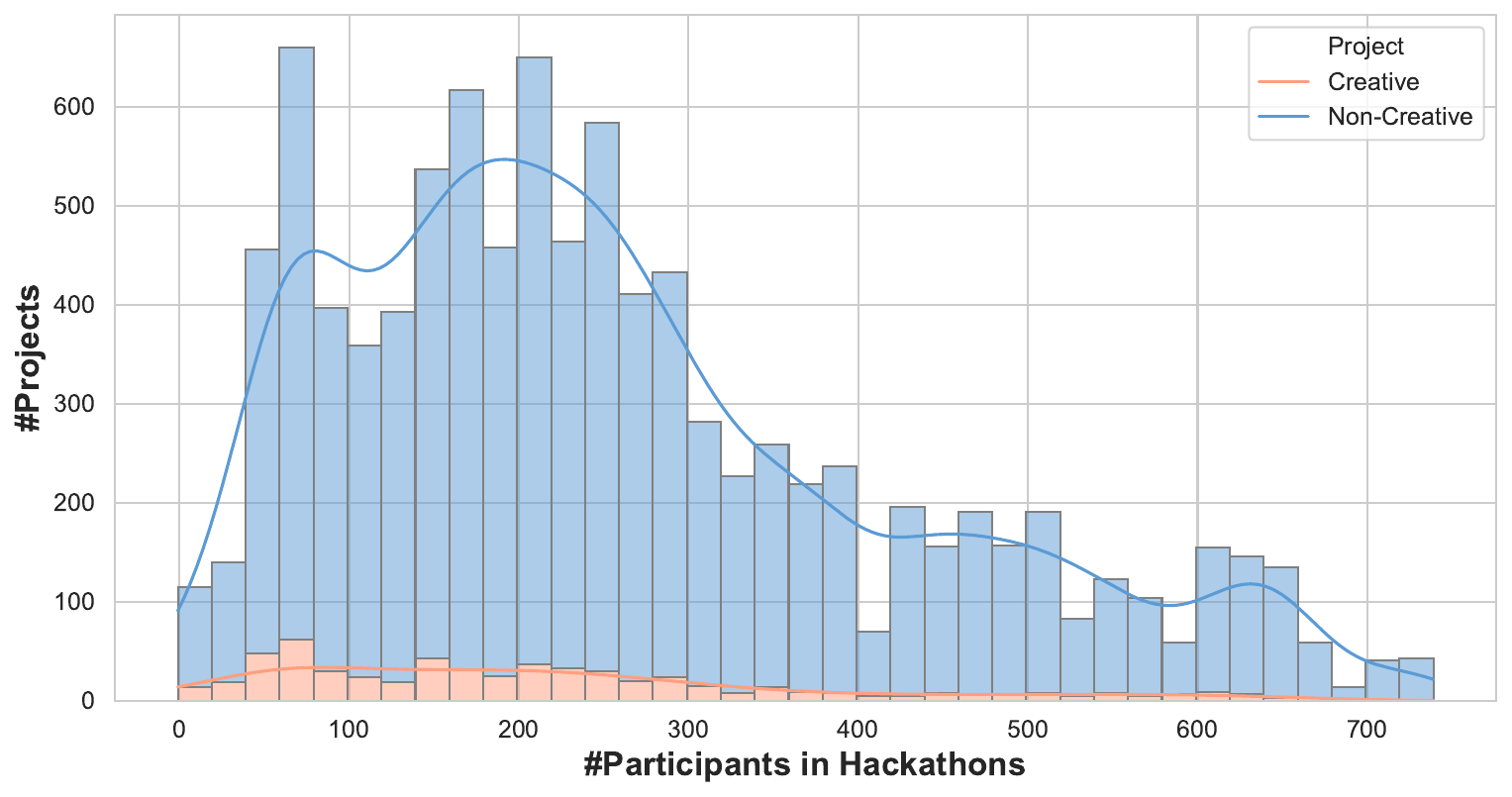}
    \caption{Top: The percentage of Creative Projects of the number of Teams by the number of Team Members. Bottom: The distribution of creative and non-creative projects by the number of participants in hackathons (outliers removed outside the 75\% percentile).}
    \Description{(Top plot) The visualization shows the relationship between team size and two key metrics in hackathons: the number of teams (shown as blue bars) and the percentage of creative projects (shown as an orange line) across teams of 1-8 members. The number of teams (blue bars) shows a clear trend: it peaks at 4-member teams with 3,365 teams, followed by 3-member teams (2,484), 2-member teams (1,965), and single-member teams (1,815). There's a sharp decline after 4 members, with only 541 five-member teams, 104 six-member teams, 25 seven-member teams, and 22 eight-member teams. The creative projects percentage (orange line) shows an interesting pattern. It starts at around 6.2\% for single-member teams, dips to approximately 4.8\% for two-member teams, and then shows an upward trend until reaching about 8.7\% for five-member teams. There's a notable drop to about 5.8\% for six-member teams, followed by a dramatic dip to around 4\% for seven-member teams. Interestingly, it spikes significantly for eight-member teams, reaching approximately 9\%. This dual visualization effectively illustrates that while 4-member teams are the most common format, the highest creativity rates are observed in very large teams (8 members) and moderately large teams (5 members), though these larger team sizes are much less frequent in practice.
    (Bottom plot) This plot illustrates the distribution of hackathon participation, comparing "Creative" and "Non-Creative" hackathons through a histogram overlaid with density curves. The x-axis shows the number of participants (ranging from 0 to 700), while the y-axis displays the frequency of hackathons (ranging from 0 to approximately 600). The data reveals that Non-Creative hackathons (shown in blue) significantly dominate both in frequency and participation rates, with a notable peak around 200-300 participants and the highest frequency bars reaching about 600 hackathons. In contrast, Creative hackathons (shown in pink/red) maintain consistently lower frequencies throughout and are concentrated in the lower participant range. The overall right-skewed distribution indicates that most hackathons tend to have fewer than 400 participants, with declining frequency for larger participant counts, suggesting that non-creative or project-based hackathons are both more common and typically attract larger participant pools compared to creative hackathons.
    }
    \label{fig:team_size_creative_project}
\end{figure}

\section{How are Hackathons Creative?}
\label{findings}
In this section, we focus on the findings which tell us something about the factors which seem to have a significant effect for whether a hackathon project is considered creative.

%\subsection{How are Hackathons Creative?}
\label{How are Hackathons Creative?}
We fitted a mixed-effects logistic regression model to the processed data (c.f. Section~\ref{sec:setup}) in order to investigate the relationship between a project being considered creative – operationalized as it being among the winning projects of a hackathon and with unusual combinations of imported packages in GitHub repositories - and the aforementioned aspects that relate to individual, team, and hackathon characteristics. We model creativity as a dichotomous variable and consequently used logistic regression for our analysis. We modeled hackathon-related aspects as a random effect because projects that take place at the same hackathon are not independent. The results of the logistic regression are presented in Table \ref{tab:regression}. We only included variables in the model that were significantly associated with creative projects.

Our findings show effects related to individual, team, project, and event characteristics. Related to event characteristics we found that hackathon size is negatively associated with creativity ($\beta = -0.226608$, $p < 0.05$), suggesting that smaller hackathons tend to be more conducive to creative outcomes. 
Further analyzing this aspect we created a plot that shows creative projects compared to the size of a hackathon (Fig.~\ref{fig:team_size_creative_project}, right). Due to the long-tail distribution of our dataset -- the largest hackathon had 11553 participants -- we only plotted the bottom 75\% of events. This plot shows that most creative projects were conducted in hackathons that had around 60 to 80 participants.
% The biggest hackathon includes 11553 participants, while the smallest one has only 4 participants, on average there are 401 participants in a hackathon. 
%This could be explained by smaller hackathons providing an environment where participants have more opportunities to interact and collaborate.
%Moreover, participants might have better access to resources such as mentors or tools in smaller events which in turn can enhance their ability to develop creative solutions.

Related to team characteristics we found that teams facing a lot of competition -- as in many teams vying for few prizes -- to be a strong predictor of creativity ($\beta = 0.424168$, $p < 0.001$) indicating that perceived competition might foster creativity. 
In addition, we also found team size as in the number of team members to be positively associated with creativity ($\beta = 0.156271$, $p < 0.05$). Further analyzing this aspect we again created a plot that shows creative projects compared to the number of team members (Fig.~\ref{fig:team_size_creative_project}, left). The plot only shows teams from 1 to 8 members because none of the 42 projects that had more than 8 members in our dataset were recognized as being creative. The plot shows that teams with 4, 5, and 8 members -- which could be an outlier due to the small number of teams that had 8 members -- exhibit more creative projects.
%Of a total of 10,363 projects, there are only 42 projects that have more than 8 team members, and none of which are recognized as creative projects.
%As shown in Fig.~\ref{fig:team_size_creative_project}(Left), the percentage of creative projects steadily increases with team sizes ranging from two to five members. Although projects with eight members are uncommon, they exhibit a notably high percentage of creative projects.

%This finding, however, goes counter to related works on general team creativity who found that larger teams may result in fewer ideas being offered due to evaluation apprehension, see e.g. \cite{byron2023buildingblocks}.
%A possible explanation may be that teams need to not only create ideas but also produce a functioning prototype in hackathons.
%This appears reasonable as larger teams make for a larger possibility of creative ideas to emerge. 

At the same time, our analysis showed a negative association between the number of interests in a team and creativity ($\beta = -0.305535$, $p < 0.05$). 
%This, again, appears counterintuitive because diversity of viewpoints is commonly associated with creativity \cite{hundschell2022diversityoncreativity}.
%In hackathons, developing a creative idea is only the start, though. 
%Teams also need to develop an artifact and diverse teams might have a harder time to quickly decide which project to work on leaving less time for development. 
%This aligns with observations from Irani's ethnographic study of a hackathon ``When there isn’t time, you don’t want to bring people into the room who are too different from you, who see things differently, or you think might create conflict.'' \cite{irani2015hackathons}.

Summarizing these findings, it appears that larger teams with coherent viewpoints that have to compete for few prizes account for more creative projects in hackathons.

Related to the projects that teams worked on, we found that the number of likes that a project received ($\beta = 0.275278$, $p < 0.1$), the number of of GitHub Watches ($\beta = 1.584174$, $p < 0.05$) and the size of GitHub projects ($\beta = 0.097405$, $p < 0.01$) -- as in file size -- to be positively associated with creativity. This finding is difficult to interpret, though, since likes might have been used to determine a winning team and a project might have received likes and GitHub watches after a team won a prize. Moreover, the size of a project being positively associated with creativity could be attributed to us utilizing unusual combinations as an indicator for creativity with larger projects having more potential for such unusual combinations. It is, however, also possible that projects with more elaborate artifacts are considered to be more creative in the context of hackathons.

Most individual aspects we considered for our model were associated with creativity. The largest positive individual predictor was the average number of hackathons an individual has won ($\beta = 0.918309$, $p < 0.001$). This appears reasonable, since we utilized winning as one of the criteria to assess creativity and it can be expected that someone who has won an event would know what it takes to win again. Just utilizing the same project again, however, does not appear to be perceived as creative as shown by the negative association between project repetition and creativity ($\beta = -0.247321$, $p < 0.05$). 
Our findings also show a positive relationship between repeated hackathon participation with the same team and creativity ($\beta = 0.484675$, $p < 0.001$). Conversely, frequent individual participation across multiple hackathons is negatively associated with creativity ($\beta = -0.7695$, $p < 0.001$). These results highlight the value of shared experiences, even if participants do not work on the same project, and suggest that common exposure through repeated events may enhance collective ideation.
%Moreover, our findings revealed a positive connection between repeated hackathon participation -- i.e. individuals participating in multiple hackathons with the same people -- and creativity ($\beta = 0.484675$, $p < 0.001$). At the same time, we found that repeated participation in general as represented by the average number of hackathons an individual participated in is negatively associated with creativity ($\beta = -0.7695$, $p < 0.001$). These findings point to the importance of joint experiences even if individuals might not have worked together on the same project. Moreover, they indicate that exposure in the form of repeated hackathon participation might foster joint ideation beyond individual events. 

We also found a negative association between perceived individual competition and creativity ($\beta = -0.220472$, $p < 0.05$) which appears surprising since competition was the largest positive predictor for creativity on a team level. One possible explanation is that larger teams might aid the confidence of individuals thus making the competition less daunting. Finally, our findings also revealed that an individual having followers on Devpost is positively associated with creativity ($\beta = 0.246149$, $p < 0.001$). This -- like the connection between a team receiving likes and creativity as discussed before -- is difficult to interpret, though, since it is likely that an individual might have gained followers after they had won.

%The average experience of team members also emerged as a significant predictor of creativity ($\beta = 0.1765$, $p < 0.05$). Teams composed of members with more hackathon experience are more likely to produce winning projects. This aligns with the suggestion that experienced participants are better equipped to navigate the challenges of a hackathon environment, including time constraints, collaboration, and prototyping.

%Another significant predictor of creativity is the number of likes a project receives on DevPost ($\beta = 0.6778$, $p < 0.001$). Projects that engage the community and receive recognition in the form of likes are more likely to be considered creative. This finding highlights the role of community validation in the creative process.

With respect to the goodness of fitting the model, the high $R^2_m$ of $0.967$ suggests that a significant proportion of the variance is explained by the fixed effects alone. Meanwhile, the $R^2_c$ of $0.970$ reflects the proportion of variance explained by both the fixed and random effects combined. Overall, the model demonstrates a strong fit to the data.

\begin{figure*}[t]
    \centering
    \includegraphics[width=\linewidth]{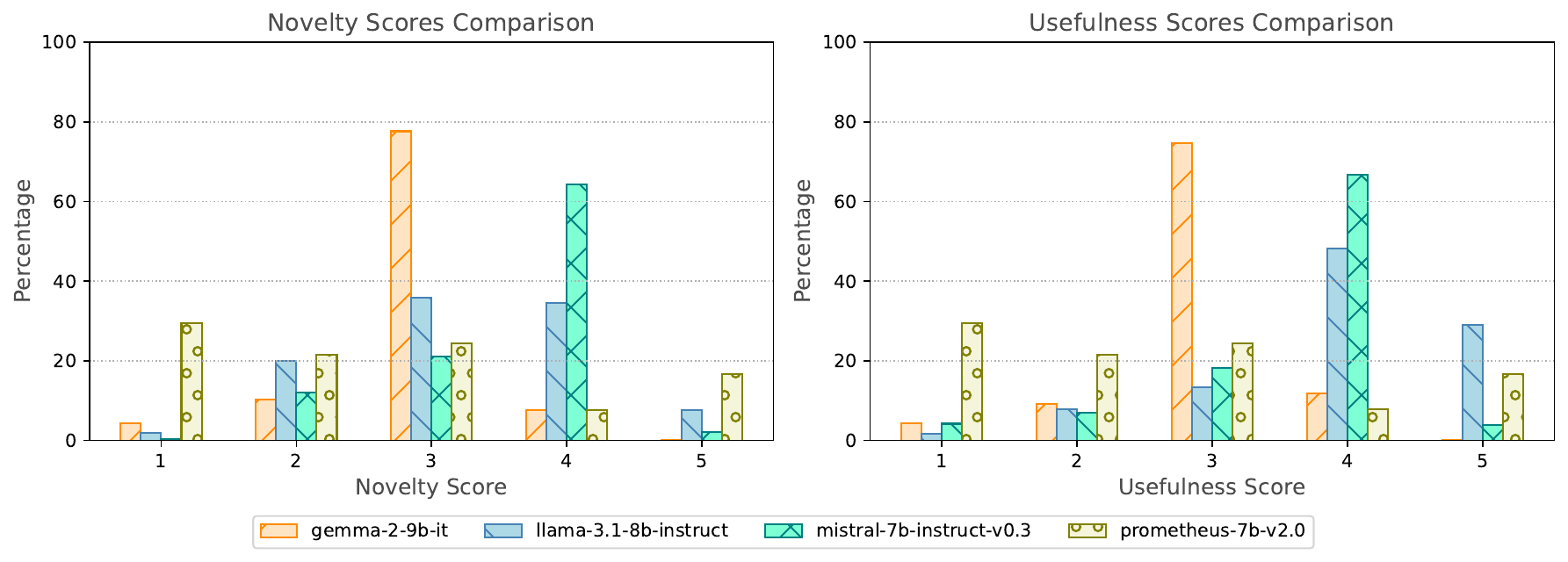}
    \caption{Score distribution of Novelty and Usefulness scores of LLM-as-a-judge on the subset of hackathons.}
    \Description{Score distribution of Novelty and Usefulness scores of LLM-as-a-judge on the subset of hackathons. The figure shows two bar charts side by side. The left chart compares Novelty Scores, while the right chart compares Usefulness Scores, both on a scale of 1 to 5. Four LLM models are compared: gemma-2-9b-it, Llama-3.1-8b-instruct, mistral-7b-instruct-v0.3, and prometheus-7b-v2.0. For Novelty, Gemma peaks at score 3, Llama is more evenly distributed, Mistral peaks at score 4, and Prometheus is relatively uniform. For Usefulness, Gemma again peaks at 3, Llama slightly favors higher scores, Mistral shows a strong preference for score 4, and Prometheus is more evenly distributed. The charts reveal distinct patterns in how different LLMs evaluate novelty and usefulness of hackathon projects.}
    \label{fig:score-distribution-llm-as-judge}
\end{figure*}

\section{Exploring LLM-as-a-judge}
\label{exploring llm as-a-judge}

We investigate LLM-based methods, inspired by \cite{luchini2023automatic} and others, to expand our exploration of creativity evaluation at scale. Specifically, we employ four LLMs to evaluate the usefulness and novelty of project descriptions. The models we use for judging the hackathon descriptions are \textsc{Llama-3.1-8b-instruct}~\cite{dubey2024llama}, \textsc{Mistral-7b-instruct-v0.3}~\cite{jiang2023mistral}, \textsc{Gemma-2-9b-it}~\cite{team2024gemma}, and \textsc{Prometheus-7b-v2.0}~\cite{kim2024prometheus}. We selected these models for two main reasons. First, they are all instruction-following models, meaning users can prompt them to perform specific tasks. Second, their smaller size fits our computational constraints, and avoids incurring an excessive environmental impact. There is no direct difference between the models, apart from the pre-training data or instruction tuning data that the organizations that trained the models do \textit{not} release. Therefore, we also account for diversity of models. In the case of \textsc{Prometheus-7b-v2.0}, this model is specifically trained to score in rubrics and give feedback, which should be well-suited for our experiments. 
%We presume this is relevant for our task of rating hackathon descriptions.
%We plan to explore judging text with larger models in future work. 
Figure~\ref{fig:prompt} in Appendix~\ref{app:prompt} presents the detailed prompt we use. We base our descriptive prompt on~\cite{kim2024prometheus}. 
Both \cite{kim2023prometheus} and \cite{kim2024prometheus} advocate for a detailed prompt that specifies what to evaluate in a text, rather than simply asking to ``rate this text 1 to 5''.

In Figure~\ref{fig:score-distribution-llm-as-judge}, we show a barplot of the novelty and usefulness scores from the four LLM-as-a-judges on a subsample of 21,318 hackathon descriptions (randomly from the total 193,353 descriptions) of both creative and non-creative projects. %In several cases, we removed descriptions where the model did not give a score, 0.03\% for \textsc{Gemma-2-9b-it}, 5.3\% for \textsc{Llama-3.1-8b-instruct}, 12.0\% for \textsc{Mistral-7b-instruct-v0.3}, and 40.6\% for \textsc{Prometheus-7b-v2.0}. We did not attempt to regenerate the output of the models.
The figure reveals distinct patterns. 
For example, we notice that \textsc{Gemma-2-9b-it} consistently favors median scores, with peaks at score 3 for both novelty ($\pm$80\%) and usefulness ($\pm$75\%). 
\textsc{Mistral-7b-instruct-v0.3} and \textsc{Llama-3.1-8b-instruct} show a tendency towards higher ratings, both peaking at score 4 ($\pm$65\%) for both metrics. 
\textsc{Prometheus-7b-v2.0} displays the most uniform distribution across scores 2-5. Usefulness scores generally exceed novelty scores, particularly for higher ratings. 

This suggests, according to LLMs, hackathon participants often create practical solutions, even if not always novel. 
The variation between models highlights evaluation subjectivity and the need for multiple judges. 
Most entries receive middle to upper-middle scores, indicating room for improvement in both novelty and usefulness. 
The data shows a correlation between novelty and usefulness scores across all models, suggesting these attributes often coincide in hackathon entries.

\subsection{Comparing Humans and LLM-as-a-judge Ratings}\label{subsec:llmasjudge}

\begin{figure*}[hpbt]
\centering
\begin{subfigure}{0.35\textwidth}
    \includegraphics[width=\textwidth]{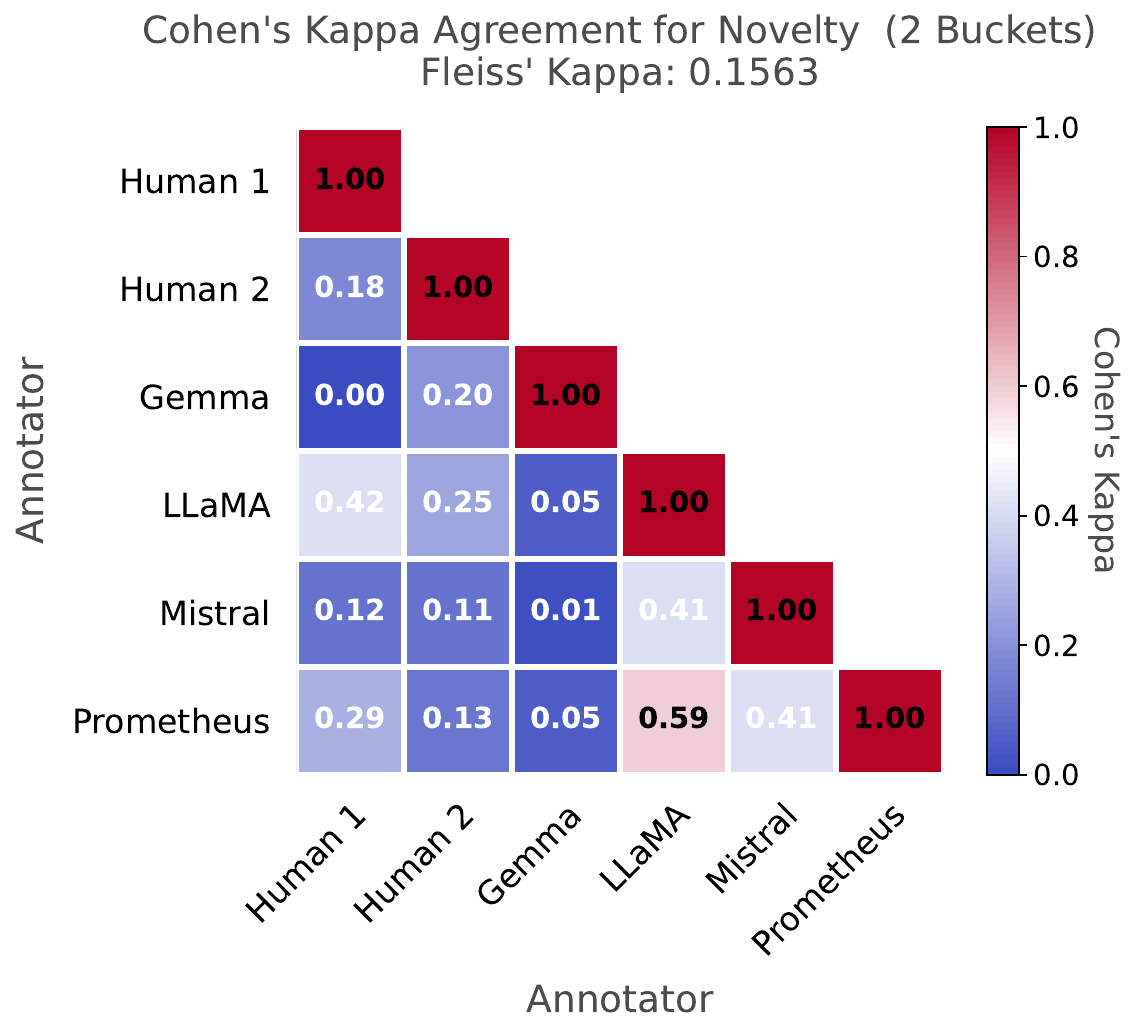}
    \caption{Inter-rater reliability measured by Cohen's Kappa for Novelty. The value among all raters is 0.1563.}
    \Description{(a) Inter-rater reliability measured by Cohen's Kappa for Novelty. The value among all raters is 0.1563.}
    \label{fig:heatmap-novelty-(1-3)-(4-5)}
\end{subfigure}
\hspace{1em}
\begin{subfigure}{0.35\textwidth}
    \includegraphics[width=\textwidth]{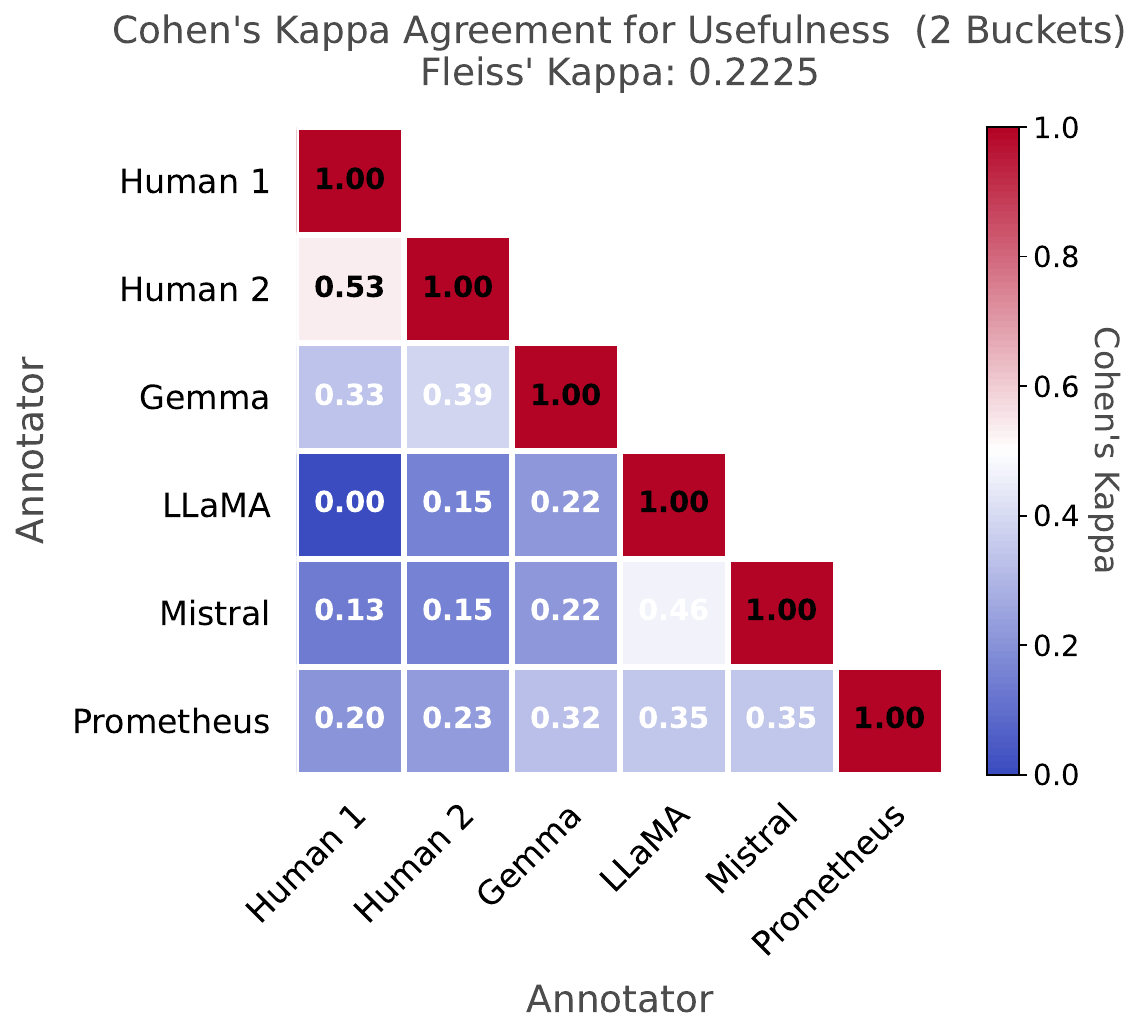}
    \caption{Inter-rater reliability measured by Cohen's Kappa for Usefulness. The value among all raters is 0.2225.}
    \Description{(b) Inter-rater reliability measured by Cohen's Kappa for Usefulness. The value among all raters is 0.2225.}
    \label{fig:heatmap-usefulness-(1-3)-(4-5)}
\end{subfigure}

\begin{subfigure}{0.35\textwidth}
    \includegraphics[width=\textwidth]{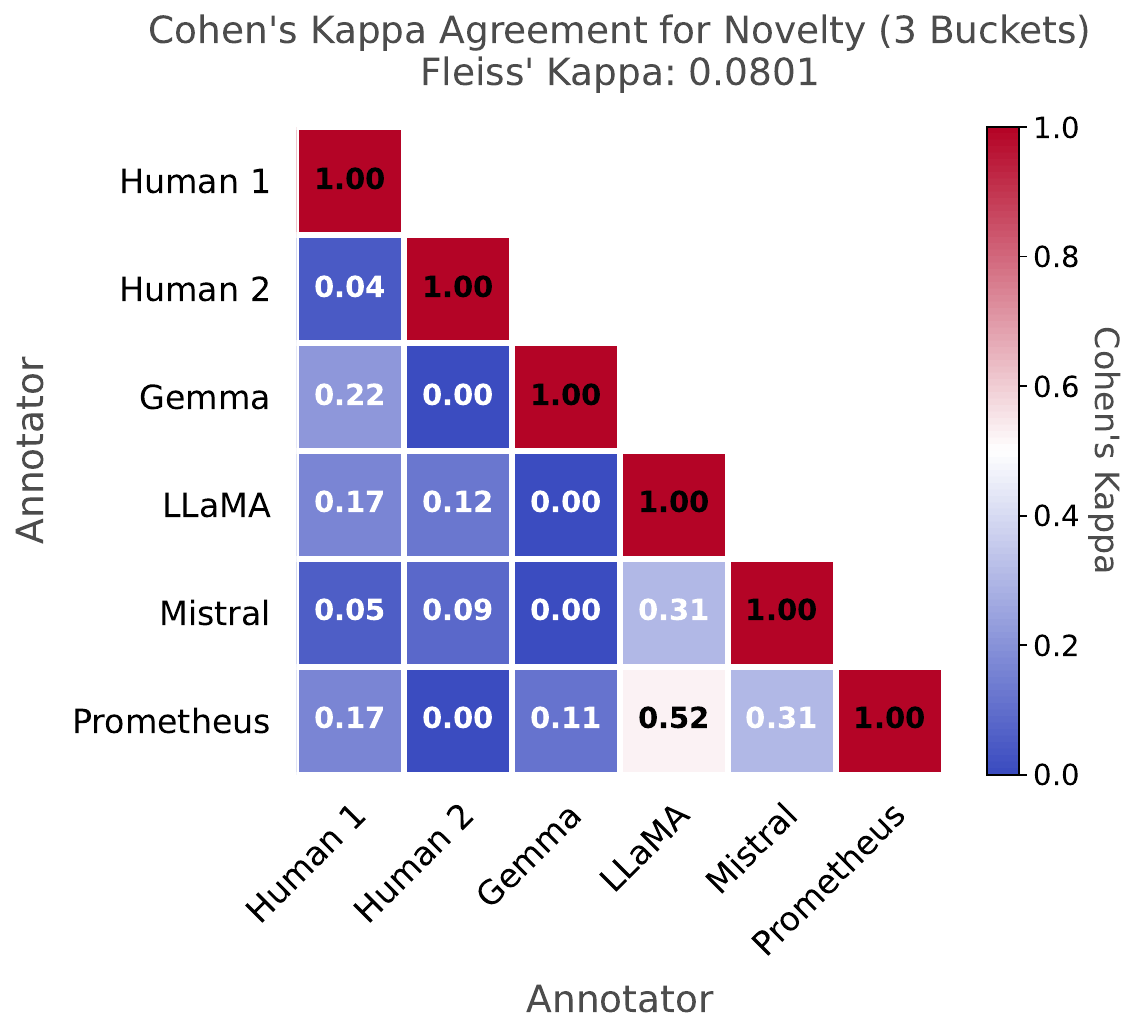}
    \caption{Inter-rater reliability measured by Cohen's Kappa for Novelty. The value among all raters is 0.0801.}
    \Description{(c) Inter-rater reliability measured by Cohen's Kappa for Novelty. The value among all raters is 0.0801.}
    \label{fig:heatmap-novelty-(1-2)-3-(4-5)}
\end{subfigure}
\hspace{1em}
\begin{subfigure}{0.35\textwidth}
    \includegraphics[width=\textwidth]{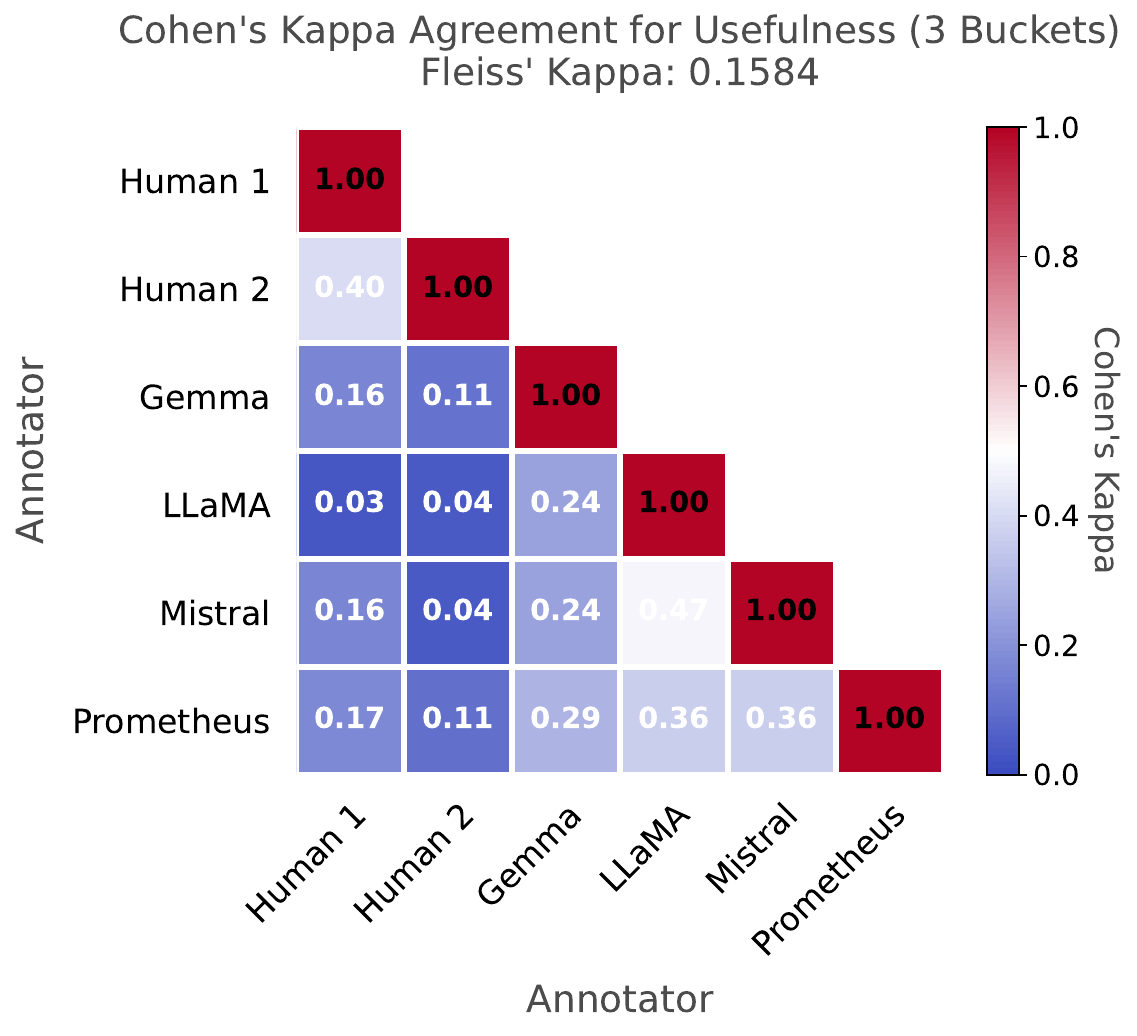}
    \caption{Inter-rater reliability measured by Cohen's Kappa for Usefulness. The value among all raters is 0.1584.}
    \Description{(d) Inter-rater reliability measured by Cohen's Kappa for Usefulness. The value among all raters is 0.1584.}
    \label{fig:heatmap-usefulness-(1-2)-3-(4-5)}
\end{subfigure}

\begin{subfigure}{0.4\textwidth}
    \includegraphics[width=\textwidth]{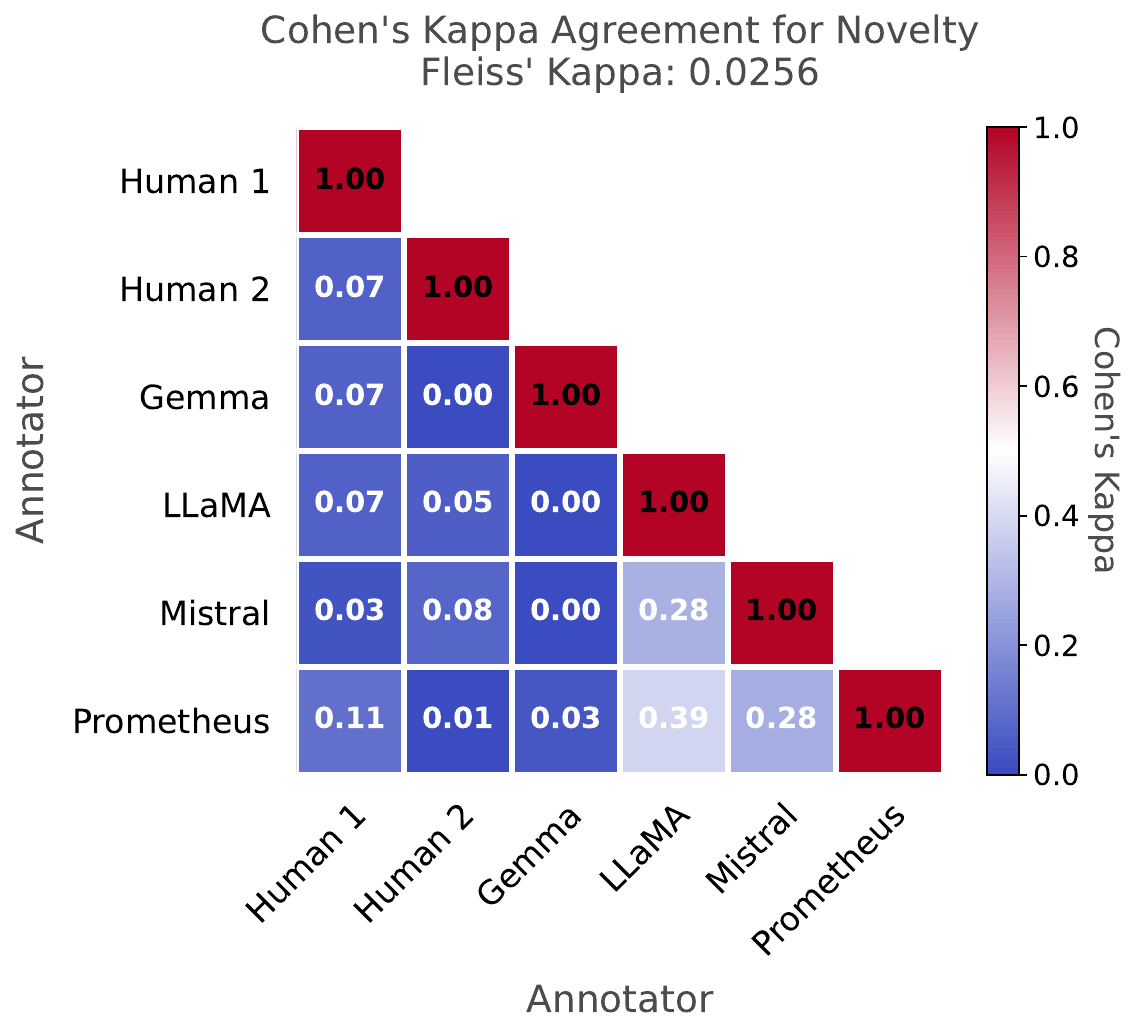}
    \caption{Inter-rater reliability measured by Cohen's Kappa for Novelty. The value among all raters is 0.0256.}
    \Description{(e) Inter-rater reliability measured by Cohen's Kappa for Novelty. The value among all raters is 0.0256.}
    \label{fig:heatmap-novelty-(1-5)}
\end{subfigure}
\hspace{1em}
\begin{subfigure}{0.4\textwidth}
    \includegraphics[width=\textwidth]{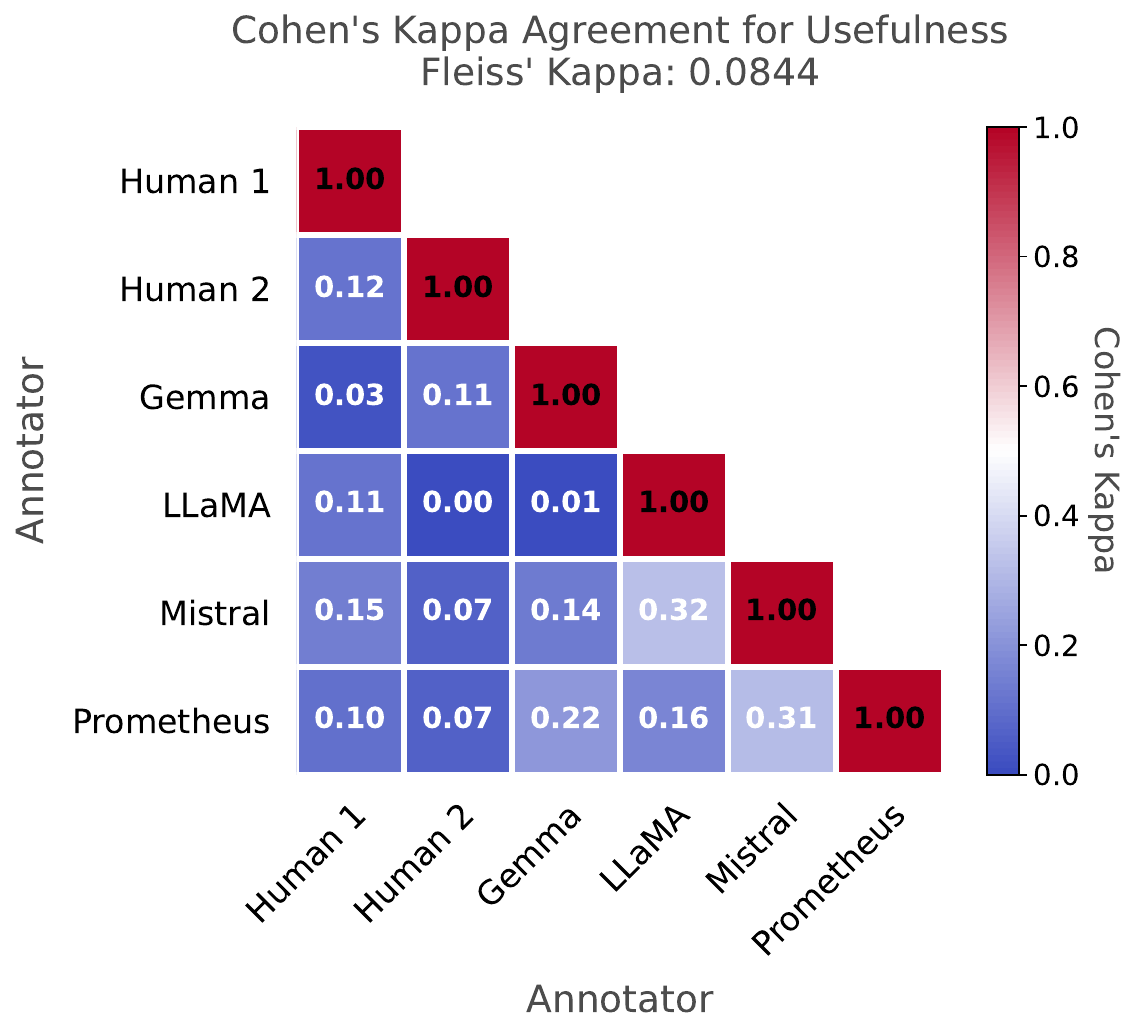}
    \caption{Inter-rater reliability measured by Cohen's Kappa for Usefulness. The value among all raters is 0.0844.}
    \Description{(f) Inter-rater reliability measured by Cohen's Kappa for Usefulness. The value among all raters is 0.0844.}
    \label{fig:heatmap-usefulness-(1-5)}
\end{subfigure}

\caption{Heatmaps of inter-rater reliability for both novelty and usefulness in hackathons in different orders of granularity, i.e., how the scores are bucketed. (a-b) contain scores (1-2-3, 4-5), (c-d) contains scores (1-2, 3, 4-5), and (e-f) contains (1-5) in separate buckets.}
\Description{Heatmaps of inter-rater reliability for both novelty and usefulness in hackathons. (a) Inter-rater reliability measured by Cohen's Kappa for Novelty. The overall inter-rater reliability among all raters is indicated by Fleiss' Kappa: 0.1563 for two buckets. (b) Inter-rater reliability measured by Cohen's Kappa for Usefulness. The overall inter-rater reliability among all raters is indicated by Fleiss' Kappa: 0.2225 for two buckets. 
(c) Inter-rater reliability measured by Cohen's Kappa for Novelty. The overall inter-rater reliability among all raters is indicated by Fleiss' Kappa: 0.0801 for three buckets. (d) Inter-rater reliability measured by Cohen's Kappa for Usefulness. The overall inter-rater reliability among all raters is indicated by Fleiss' Kappa: 0.1584 for three buckets
(e) Inter-rater reliability measured by Cohen's Kappa for Novelty. The overall inter-rater reliability among all raters is indicated by Fleiss' Kappa: 0.0256 for single buckets. (f) Inter-rater reliability measured by Cohen's Kappa for Usefulness. The overall inter-rater reliability among all raters is indicated by Fleiss' Kappa: 0.0844 for single buckets
Each cell shows the Cohen's Kappa value between two raters, with Human 1, Human 2, Gemma, Llama, Mistral, and Prometheus as raters. The heatmaps use a color scale from blue (low agreement) to red (high agreement), with diagonal values of 1.00 representing perfect self-agreement.}   

\label{fig:heatmaps-agreement}
\end{figure*}
As a next step in exploring LLMs-as-a-judge to evaluate creativity, we wanted to investigate the agreement between humans and LLMs evaluating the novelty and usefulness of hackathon projects.
We randomly sampled 30 projects and had two human raters---experienced with hackathon organization, participation, and judging---evaluate each project. 
In turn, these evaluations are then compared to the novelty and usefulness scores from the LLMs~\ref{subsec:llmasjudge}.

We measured the inter-rater reliability with unweighted Cohen's $\kappa$ ~\cite{cohen1960coefficient}. Figures~\ref{fig:heatmaps-agreement} to~\ref{fig:heatmap-usefulness-(1-5)} display these results, with three levels of granularity: (a-b) scores bucketed as 1–3 and 4–5, (c-d) bucketed as 1–2, 3, and 4–5, and (e-f) each score treated as its own bucket. 
The human raters showed low agreement on novelty, with Cohen's Kappa decreasing as granularity increased (e.g., $k = 0.18$ in a-b vs. $k = 0.04$ in e-f), suggesting subjectivity in evaluating novelty. 
Conversely, they showed higher consistency for usefulness ($k = 0.53$ in a-b, $k = 0.12$ in e-f), indicating shared criteria for this dimension. 
This aligns with related work, which indicate that when rating creativity in the form of novelty and usefulness in complex projects (like urban planning), judges put more emphasis on usefulness when rating such projects whereas they put more emphasis on novelty in alternative uses tests \cite{lloydcox2022evaluatingcreativity}.

Agreement between humans and LLMs varied. Human 1 agreed more with Llama ($k = 0.42$ in a-b), and both humans showed fair agreement with Gemma. Among LLMs, Prometheus and Llama achieved the highest agreement for novelty ($k = 0.59$ in a-b), while Mistral and Llama had the best agreement for usefulness ($k = 0.46$). 
Fleiss’ Kappa scores further indicated stronger agreement for usefulness than novelty across all settings.
The discrepancies among LLMs likely stem from differences in pre-training data, as models are trained on varying data snapshots. 

In creativity evaluations, it is common to prioritize rater consistency—how raters rate items relative to each other—over achieving absolute agreement, where all raters provide identical scores \cite{ceh2022assessingraters}.
Therefore, it is not surprising that agreement between models and humans is low. 
Hence, instead of framing LLMs as an equal to human creativity raters, we propose Human-AI collaboration on creativity evaluation. 

\subsection{Towards AI Collaborative Creativity Evaluations}
\label{Towards AI collaborative creativity evaluations}
The same two human raters also had a detailed discussion on the generated output of the four judge models. Empirically, the raters decided on two additional ratings for the output of the LLMs, one is about \textit{representativeness} (R; rating 1--5, i.e., how well the LLM output encapsulated the actual hackathon description?) and \textit{expert usefulness} (EU; rating 1--5, i.e., can an expert or rater make use of this generated output in their decision making?). We bucket the scores 1--2 and 3--5 and compare agreement over these. For Prometheus, there is high agreement among the human raters that the generated output is both representative and useful for an expert to use to integrate in their decision (R; $k$ = 0.47, EU; $k$ = 0.44). For the other models, we observed low agreement among representativeness, likely due to the generated output being shorter than Prometheus' (R; $0.05 < k < 0.2$). With respect to expert usefulness, we overall see a higher agreement as the generated output can still support a judge in their evaluation of a hackathon project's creativity  (EU; $0.20 < k < 0.30$).

More broadly, there are several key findings that emerged highlighting the complexities involved in using LLMs for such evaluations. 
Firstly, assessing creativity requires temporal context; a project that was novel and useful at a certain point in time may lose its novelty as similar solutions emerge over time. This necessitates situating creative assessments within the appropriate time frame and considering changing conditions, such as those brought about by the COVID-19 pandemic, which can affect a project's relevance. 
%A concrete example is that Llama3.1 rated a project concept as unique, ``\textit{combining brainwave analysis with video speed control}''. However, from one of the human annotators, they note it is not entirely new, as brain-computer interfaces exist. Still, the application to video speed control is an original approach.

Secondly, the models often provided overly optimistic evaluations with little nuance (i.e., scoring similarly), sometimes based on unfounded assumptions about the projects' functionalities. They occasionally perceived usefulness where human evaluators did not, leading to discrepancies. For instance, Prometheus and Llama disagreed on how much a project's approach differed from existing ones, likely due to variations in their training data and embedded knowledge. This inconsistency poses a challenge for human evaluators, especially novices, who may not have the expertise to verify the models' claims. 
For example, Llama3.1 generated ``\textit{The project description lacks a clear and unique concept, relying heavily on existing ideas (e.g., local hack day inspiration). However, it addresses a practical problem (communication assistance) and presents a feasible solution (a speaking assistant). The use of Python as a development tool is common, but the potential for adding features to the bot suggests some originality in its approach}''. This however, is based on an almost empty project description.

Finally, the models tended to focus on evaluating the prototypes rather than the underlying ideas or their potential for future development. This raises questions about the models' ability to assess the possible potential of creative projects.
Moreover, hackathons are celebrated not only for fostering the development of creative ideas but also for offering participants valuable educational and hands-on experiences.
In one example from a project description, the creators described how ``Neither of us have ever done anything with Machine Learning, so this was something we were proud of.''

%fully appreciate the breadth of outcomes from such events. 
% \textbf{TODO add concrete examples}

\section{Discussion}
According to Devpost\footnote{\href{https://devpost.com/hackathons}{https://devpost.com/hackathons}}, the largest hackathon database, over 1,000 hackathons are held annually, though the actual number is likely higher, as many are not registered \cite{falk2024future}.
From a creativity research perspective, hackathons offer valuable opportunities to study large-scale, real-world projects addressing problems creatively.
This paper is a first exploration of evaluating creativity at scale. Prior work has mainly focused on studying few events that took place in a specific context and that were organized and attended by specific individuals with certain backgrounds, motivations and goals~\cite{gama2023developers,rosell2014internalhackathons,Briscoe2014DigitalIT,frey2016innovationdrivenHackathon} which limits the usefulness of the reported findings to the studied contexts. To expand this body of knowledge we conducted an analysis on the dataset of hackathon projects, repeating and expanding on the approach by Fang, Herbsleb and Vasilescu \cite{fang2024novelty}, and subsequently exploring automated evaluation of creativity by using LLMs-as-a-judge, inspired by the approach by \cite{luchini2023automatic}. 
In addition to insights on organizing hackathons to foster creativity (see section \ref{How are Hackathons Creative?}), we discuss our evaluation method, and suggest directions for future research. 

Plucker, Beghetto and Dow emphasized the importance of considering ``creativity for whom'' and ``creativity in what context'' \cite{plucker2004standarddefinitioncreativity} when defining creativity.
Creativity is not a uniform but operates on different scales. 
Kaufman and Beghetto's Four C Model of Creativity aims to capture these scales: mini-C (i.e. personal and developmental aspects of creativity inhering in learning processes), Little-C (i.e. everyday creativity which the average person may engage in), Pro-C (i.e. highly accomplished but not yet eminent forms of creative expression), and Big-C (i.e. creative genius or extraordinary creativity) \cite{kaufman2009BeyondBigandLittle}.
Even though a hackathon Creation may not have been identified as creative in our analysis, the Creators may have experienced it as very creative to them personally, indicating a mini-C experience of creativity, which we also observed as the last point in section \ref{Towards AI collaborative creativity evaluations}.  These mini-C experiences, including personal learning, are key motivators for hackathon participation \cite{simonofski2020participationinhackathons}.
Future evaluation metrics could consider multiple levels of creativity to account for \textit{tangible} (project-level) and \textit{intangible} (experience and learning) outcomes \cite{falk2021whatdo}.
%One framework for discerning different levels of creativity -- such as creative ideas that are groundbreaking vs. an individual learning about already-existing creative ideas for the first time -- is the Four C-Model of Creativity \cite{kaufman2009BeyondBigandLittle}.
%Considering this model, hackathons can foster both mini-c creativity (i.e. creators' own personal growth and learning), `little-c' creativity (everyday problem-solving), pro-C creativity (``the developmental and effortful progression beyond little-c'' \cite{kaufman2009BeyondBigandLittle}) and potentially `big-C' creativity (groundbreaking innovations) \cite{kaufman2009BeyondBigandLittle}. 

\subsection{Take-Aways for Organizers}
One contribution of this work is to develop our understanding of how individual participants, teams, and event characteristics foster creativity in hackathons (RQ2), leading to concrete recommendations for hackathon organizers. Moreover, we expect our findings to be useful for organizers of hackathons across different contexts, since they were obtained from a large-scale study rather than a study of few specific events.

\textbf{(1) Creative projects are more likely to emerge in hackathons with around 60 to 80 participants} (see fig. \ref{fig:team_size_creative_project}), potentially due to greater opportunities for team interaction, progress presentations, and feedback. Smaller events also may offer more to resources such as mentors or tools enhancing teams' ability to develop creative solutions.
These findings add nuance to Attalah, Nylund and Brem's discussion on how hackathons is a form of collective creativity which can feed into collective intelligence \cite{attalah2023whoCapturesValueFromHackathons}. Our results suggest collective creativity is not universal across \textit{all} hackathons but may depend on event size, with smaller events being more conducive than larger ones. 
This is related to our finding that \textbf{repeated collective participation in different hackathons with the same team is related to creative projects} contrary to repeated individual participation in hackathons.

\textbf{(2) Larger team sizes are positively associated with creativity,} countering prior research suggesting that larger teams inhibit idea generation due to evaluation apprehension  \cite{byron2023buildingblocks}. In hackathons, teams must both generate ideas and produce functioning prototypes. This dual requirement may explain why larger teams succeeded despite the risk of apprehension. Our analysis suggests an optimal team size of four to five members may represent this sweet spot (Fig.~\ref{fig:team_size_creative_project}, left), however future research is needed to validate this finding.

\textbf{(3) Perceived competition can foster creativity on a team level but can at the same time be detrimental on an individual level}. Organizers thus need to walk the tight rope of making the event competitive while at the same time providing an environment where individuals feel safe enough to develop risky and creative ideas \cite{gross2020creativityunderfire}. One approach could be to provide a few prizes for the most creative projects (fostering perceived team competition), while at the same time making sure that individuals work in larger (see the next point) stable teams which may shield them from perceiving the competition as too daunting.

\textbf{(4) Teams with less diverse interests are related to creative projects,} a counterintuitive finding given that functional diversity (i.e, education, job-relevant knowledge, skills, and abilities) often support creativity \cite{reiter-palmon2012teamcreativity}.
In hackathons, developing a creative idea is only the start, though. 
Teams also need to develop an artifact and diverse teams might have a harder time to quickly decide which project to work on leaving less time for development. 
This aligns with observations from Irani's ethnographic study of a hackathon ``When there isn’t time, you don’t want to bring people into the room who are too different from you, who see things differently, or you think might create conflict.'' \cite{irani2015hackathons}.

\textbf{(5) Prior hackathon experience is positively associated with creative projects}, potentially because ideas evolve over time. Experienced participants may draw on past hackathons to refine their projects and apply domain knowledge effectively \cite{Tenorio2019RethinkingSO}. While prior research highlights how diverse teams with various knowledge bases lead to more creativity \cite{nolte2020organize, carmit2012beyondindividualcreativity}, our findings suggest that experience itself is a key driver in creativity in hackathons. 

%Potentially the folowing snippet fromt the related work could be discussed above
%``Pre-registered teams may improve teamwork but may jeopardize creativity as such teams are usually made up of participants with similar backgrounds \cite{nolte2020organize} whereas diverse teams have shown increased creativity \cite{carmit2012beyondindividualcreativity}.''

Finally, \textbf{(6) creative projects receive more likes on Devpost than less creative ones}, suggesting that other participants are skilled at identifying creativity. It is generally agreed upon that creative outcomes are best judged by experts familiar with the relevant domain, as creativity is determined by their independent agreement. Thus if the participants themselves are considered to be experts, our findings fit well with theory \cite{cseh2019scattered, amabile1982social}. 
Another explanation for our finding could also be that teams might actively promote their project if they perceive it to be creative, which might manifest itself in the form of likes in our dataset. A similar finding has been reported in the context of hackathon project continuation, where one of the predictors of the long-term survival of projects was reported to be teams promoting their project \cite{nolte2020WhatHappensToAllTheseHackathonProjects}.

These findings may be valuable for organizers to support them in fostering creative projects in their events. It might be advisable to run smaller (60 to 80 participants) rather than larger events that offer opportunities for teams to interact and see each other’s work during the event. Moreover, while the focus is often on attracting newcomers, it might be advisable to invite experienced teams with individuals who have participated in multiple hackathons. In relation to this point, it might also be advantageous to encourage people to participate in hackathons in the future. The interactions that they experienced during one event might as well be the spark that helped them develop a creative project during the next one. Finally, our findings show that it might be advisable to trust the wisdom of the hackathon peers when trying to identify creative projects.

\subsection{Evaluating Creativity at Scale: Challenges and Opportunities}
To address RQ1, we needed to take several decisions to operationalize theory from creativity research into measurable constructs for data science.
Although Plucker, Beghetto and Dow called for the creativity research community to develop a unified definition of creativity 20 years ago \cite{plucker2004standarddefinitioncreativity}, ``the field of creativity research has continued to lack solidarity and cohesiveness'' \cite{puryear2020definingCreativity}.
Needless to say, it is a challenge to then quantify and operationalize creativity constructs to enable a large scale data analysis and, furthermore, develop an automated evaluation.
While our contribution is moving large-scale evaluation towards real-world data compared to prior, related work, filtering and analyzing a large-scale dataset is a trade-off between accuracy and feasibility.
Our findings are therefore limited to the kind of method which we have constructed in order to answer our research question of how hackathons are creative, and in the following subsections we discuss the challenges and opportunities of our approach.

\subsubsection{Operationalizing Novelty and Usefulness}
As this research is an initial exploration of a large-scale data analysis and automated evaluation of creativity, we do not claim that our operationalization of the creativity definition, or its constructs of \textit{novelty} and \textit{usefulness} are the only way nor the most accurate way to identify creative hackathon projects from the dataset. 
Our definition of novelty and usefulness may have excluded some hackathon projects which others may have deemed creative.
While we aimed at replicating Fang, Herbsleb and Vasilescu's approach for operationalizing novelty \cite{fang2024novelty} -- by finding projects with unusual combinations of software packages and libraries for the five most popular programming languages -- it is just one way of capturing novelty. 
%In addition to replicating an approach to operationalizing novelty as pursued by others, we found Python to be suitable because of its popularity across different application areas.
Although we would argue that we thereby should capture a good representation of creative projects, this of course limits the analyzed dataset and potentially excludes some hackathon projects which could be considered creative. 
Furthermore, this approach also only captures creative software development, whereas creativity in hackathon projects may also happen on the user interface/interaction design side of things, where a very basic and "uncreative" combination of software is still used to develop a very novel and creative interface/interaction design. 

To operationalize usefulness in the context of a hackathon, we utilized the ``winner''-tag in our dataset, hypothesizing that this could serve as a proxy for an expert evaluating a project as having addressed a challenge or contributing to the theme of an event in a useful and creative way. This operationalization, however, has limitations in that from the dataset we used it is not clear which criteria hackathon organizers utilized to judge projects. Moreover, some of the winners might be chosen by popular vote. It is still reasonable to assume that it is more likely for teams to win at a hackathon that created an artifact which can be considered useful to address a challenge or create an opportunity that did not exist before.

We delimited our analysis of Creations to a single experiment in which we measured the cosine similarities between sentence representations of textual representations including, e.g., the requirements of hackathons.
While a more large-scale investigation could have afforded further insights into this particular aspect of hackathons, we leave it to future work to explore, e.g., using LLMs-as-a-judge to enable such large-scale analysis without being overly costly.

Future research could explore approaches for operationalizing creativity constructs further to validate them, or explore other ways of operationalizing novelty and usefulness in meaningful ways. 
Hackathon platforms such as Devpost could consider creating templates for hackathon descriptions which enable users to reflect on and write how their project is creative in terms of novelty and usefulness, to enable creativity assessment not just for creativity researchers but also for hackathon judges and perhaps recruiters as well.

\subsubsection{LLM-as-a(n additional)-Judge}
%Our exploratory study show that analyzing creativity with large-scale data and using automated assessments of creativity remains a challenge to do.
%As subsection \ref{subsec:llmasjudge} illustrated, there were some disagreements among all the raters---humans and LLMs alike---about the creativity of the 30 randomly-selected projects.
%Add argument that the goal with creativity evaluation, however, is not necessarily absolute agreement \cite{ceh2022assessingraters}

The motivation for exploring LLMs as-a-judge in large-scale creativity evaluation is compelling (RQ3).
Because of their training on huge datasets, LLMs possess broad and detailed information about many different topics and domains which they can draw on in their inference phase.
When assessing a product for its creativity, an expert would draw on their detailed knowledge about a domain to judge whether a product within that domain is creative---i.e. novel and useful within that context---or not.
Recruiting human experts for all the different topics in our hackathon project dataset would be practically infeasible, which underscores the need for exploring automated approaches such as LLMs for this.
Hence, a few LLMs should be able to judge a large dataset with an expert view, because they possess detailed information about probably all of the relevant topics in the dataset.
This could potentially also circumvent the limitations of attempts to operationalize creativity for statistical analysis, which risk being too constrained as discussed above.

For this reason, Luchini and colleagues have therefore explored LLMs to judge the creativity of creative problem-solving tasks \cite{luchini2023automatic}.
Our findings from Section \ref{exploring llm as-a-judge}, which explores LLM as-a-judge, differ somewhat from Luchini and colleagues' approach, despite overall similarities \cite{luchini2023automatic}. While they fine-tuned language models to score quality and originality in creative problem-solving tasks, demonstrating high correlation with human scoring, we contend this method has limitations. Their models, trained on task-specific texts and human ratings, naturally correlate well when applied to similar texts. However, this correlation may be more a result of the training process on homogeneous data than a true measure of creativity assessment. Our method addresses this potential bias by using an out-of-the-box LLM, not fine-tuned on the task at hand, to investigate whether its predictions still correlate with human ratings.

Despite their potential, LLMs as judges face notable challenges. Our findings reveal their tendency to provide overly optimistic and uniform evaluations, undermining their ability to offer nuanced critique or differentiate effectively between creative projects. This phenomenon aligns with the broader trend of designing AI systems to align with ``human values'' by prioritizing servility and safety \cite{cai2024antagonistic}. While this approach has commercial success, it limits the models' capacity for critical engagement. The tradeoff between "pleasing" and "provoking" interactions  highlights a fundamental challenge for AI in balancing reinforcement with critical divergence \cite{Sherson2024, 10.1145/3519026}. For hackathon evaluations, this limitation suggests the need for mechanisms that encourage models to assess more critically and avoid falling into overly affirmative patterns.

One of the most relevant works for our discussion is by Organisciak et al.~\cite{ORGANISCIAK2023101356}, who not only fine-tuned GPT-style models similar to Luchini and colleagues, but also explored a ``few-shot'' approach with GPT-4 \cite{achiam2023gpt}, presenting several task examples in the prompt instead of attempting it ``zero-shot'' like us (i.e., no task examples). Their findings show that GPT-4 achieves relatively good correlation with human ratings on Alternative Use Tests (AUT) responses ($r$ = .70). However, we argue that hackathon project ratings involve greater subjectivity, as evidenced by our results in Section \ref{exploring llm as-a-judge}. Nevertheless, following \cite{ORGANISCIAK2023101356}'s approach, we see merit in providing demonstrations to LLMs on rating hackathon project descriptions or fine-tuning them on a set of examples. We hypothesize that this method could potentially improve agreement between humans and LLMs. Additionally, research by Kim and colleagues~\cite{kim2023prometheus} suggests that providing detailed explanations for each score in a rubric improves scoring robustness, offering another avenue for improvement. An important consideration is the current context length limitation of LLMs, typically 8,192 subwords (chunks of a word) for the model sizes we used. This constraint has implications not only for potentially lengthy hackathon project descriptions but also for applications in other domains such as patent, grant funding and academic paper analysis.

Rather than positioning LLMs as equal to human judges, we propose framing them within the paradigm of hybrid intelligence \cite{Sherson2024, Mao2024, mao2023hybrid}, where LLMs complement and enhance human evaluation and decision-making processes. For instance, in the context of judging creative projects within a single hackathon, LLMs could serve as supplementary judges, providing valuable support to human judges. They might offer synthesized overviews of large datasets, highlight trends, and pose probing questions to foster deeper insights. Beyond judging, a fine-tuned LLM trained on hackathon-specific data could act as an assistant to hackathon organizers, helping uncover statistically significant relationships between factors such as team composition, participant confidence, and hackathon planning effectiveness.
Additionally, building on the statistical analysis in Section~\ref{findings} and the predictive model for creativity introduced in Section~\ref{sec:setup}, a direction for future work would be to explore how the model's outcomes align with the creativity judgments provided by LLMs.

\subsubsection{Other Venues for Exploring Real-World Large-Scale Data on Creativity}
Data on hackathon projects are one way to identify real-world data on creativity.
For creativity researchers wishing to explore similar approaches to large-scale analyses of creativity as ours, we see opportunities in a range of other disciplines for this pursuit.
Grant writing has been suggested as a form of creative writing: ``Creativity often sets apart winnable, funded grant projects from projects that are less impressive to funders in real-world settings'' \cite{Gorsevski2019GrantWA}.
Continuing the theme of creative writing, papers and patents could be considered here too. 
In a similar vein, Park, Leahey and Funk conducted a large-scale analysis on papers and patents and calculated their ``consolidating or disruptive nature'' \cite{Park2023}.
Such an analysis could also explore their creativity in terms of novelty and usefulness.
Crowdsourcing social innovation platforms, like OpenIDEO\footnote{\url{https://www.openideo.com/}}, are also interesting for exploring large scale creativity evaluations \cite{rong2023crowdsourcing}.
%How might we scale up this method --> ``Papers and patents are becoming less disruptive over time'' paper
%How to report hackathon projects to enable the assessment of their creativity? 
%Which elements should be included to enable automated assessment?

%Relate to crowdsourced idea challenges as a good use case for such an assessment method
 %Compare with patent databases?

\subsection{Environmental Impact}
We acknowledge that conducting a large-scale analysis using LLMs comes with an environmental impact. Experiments were conducted using private infrastructure in Denmark, which has a carbon efficiency of 0.115 kgCO$_2$eq/kWh in the month of August. A cumulative of 109 GPU hours (summed for all four LLMs) of computation was performed on NVIDIA A40 GPUs, which has a TDP of 300 Watts. Total emissions are estimated to be 3.76 kgCO$_2$eq. Estimations were conducted using the Machine Learning Impact calculator\footnote{Find the tool here: \url{https://mlco2.github.io/impact}.} presented in \cite{lacoste2019quantifying}.

\section{Conclusion}
In this paper, we have explored methods for evaluating creativity at scale.
We frame hackathon projects as valuable real-world data to evaluate creativity, however, while we have sought to replicate and extend prior research contributions' approaches for doing this, going from process-related creativity evaluations to evaluating creativity in real-world data is challenging. 
We replicated and extended the prior work in the following ways:
(1) we operationalized not only novelty as a creativity construct, but also usefulness as a creativity construct. 
We filtered the dataset with these two operationalized constructs and discussed the most interesting findings from this subset, which we framed as creative.
(2) We explored recent calls to explore LLM-as-a-judge to provide further insights into creativity from the full dataset, and to discuss how LLMs may be framed as an additional judge which can support human raters. 
We discuss the overall challenges and opportunities for evaluating creativity from large-scale data, which remains a challenge to do. 
Our findings also challenge some of the previous work which has been done in this area and discuss future research directions.

%%
%% The acknowledgments section is defined using the "acks" environment
%% (and NOT an unnumbered section). This ensures the proper
%% identification of the section in the article metadata, and the
%% consistent spelling of the heading.
\begin{acks}
YC and JB are funded by the Carlsberg Foundation, under the Semper Ardens: Accelerate Programme (project nr. CF21-0454). 
MZ and JB are funded by the Villum Foundation (project nr. VIL57392).
We would further like to thank the Aarhus University Research Foundation and Center for Shaping digital citizenship. % please check if the wording is okay.

\end{acks}

%%
%% The next two lines define the bibliography style to be used, and
%% the bibliography file.
\bibliographystyle{ACM-Reference-Format}
\bibliography{sample-base}

%TC:ignore
%%
%% If your work has an appendix, this is the place to put it.
\appendix

\clearpage

\section{LLM-as-a-judge prompt}\label{app:prompt}
In Figure~\ref{fig:prompt}, we indicate the full model prompt we use to query the four LLMs we use in our study.

\begin{figure}[htbp]
    
\fcolorbox{black}{white}{%
  \parbox{.95\linewidth}{
\#\#\# System Prompt \#\#\#

You are a rigorous and efficient evaluation assistant for rating creativity. Your task is to provide scores and feedback systematically. Follow the format strictly: first provide the scores for Novelty and Usefulness, then the rationale for each score in under 100 words. Ensure that feedback is clear, concise, and aligned with the rubric provided, avoiding unnecessary commentary. \newline

\#\#\# User Prompt \#\#\#

Task Description: You are tasked with evaluating a project description based on two criteria: Novelty and Usefulness, using a 5-point Likert scale. 

1. Write the scores for Novelty and Usefulness as integers between 1 and 5, strictly referring to the score rubric. 
2. Provide concise feedback (within 50 words) justifying each score. 
3. Format your output as follows: 'Novelty: [RESULT] (1-5) Usefulness: [RESULT] (1-5) Feedback: [Your feedback here]'. 
4. Do not include any other text or explanation. 

Evaluation Criteria: 
Novelty: Evaluate how unique and original the project's concept, approach, or solution is. Consider whether it introduces new ideas, methods, or perspectives that differ significantly from existing ones. 
Usefulness: Evaluate how practical and appropriate the project is in addressing its targeted problem or challenge. Consider whether it effectively solves a real-world issue or meets a specific need. \newline

Project Description to Evaluate: 
\newline
\{Example Hackathon Description\}
}
}
    \caption{\textbf{Prompt for Evaluating Hackathon Descriptions.} We show the prompt we use for all four models in our study. First we give a system prompt, in the form of a ``model-should-act-as''. Then, in subsequent paragraphs, (1) we give detailed feedback for the task itself, (2) give instructions on how to evaluate the text, (3) the hackathon description, and (4) the rubrics that should be used for the final judgment.}
    \Description{Figure 6 shows a prompt for evaluating hackathon descriptions. The image contains a text box titled ``Prompt for Assessing Projects' Creativity and Novelty''. The content begins with a description of the evaluator's role as a fair judge assistant. The task description outlines four steps: 1) Write a score between 1 and 5 based on the score rubric. 2) Write concise feedback within 50 words assessing the response quality strictly based on the given score rubric. 3) The output format should be "Feedback: (write feedback for criteria) [RESULT] (an integer number between 1 and 5)". 4) Do not generate other openings, closings, or explanations. The instruction to evaluate asks to assess a project description based on two criteria on a 5-point Likert scale: Novelty and Usefulness. Novelty is defined as how unique and original the project's concept, approach, or solution is, and if it introduces new ideas, methods, or perspectives significantly different from existing ones. Usefulness is defined as how practical and appropriate the project is in addressing the problem, situation, or challenge it targets, and if it effectively solves a real-world issue or meets a specific need. The response to evaluate is indicated as "{Example Hackathon Description}". The score rubrics for both Novelty and Usefulness use a five-point scale, where 1 is Very Unoriginal/Very Unuseful, 2 is Unoriginal/Unuseful, 3 is Neutral, 4 is Original/Useful, and 5 is Very Original/Very Useful. The caption explains that this prompt is used for all four models in the study, providing a system prompt, detailed feedback instructions, the hackathon description, and the rubrics for the final judgment.}
    \label{fig:prompt}
\end{figure}

\section{Correlation Tables}
\label{appendix:correlation}

\begin{table*}[htbp]
    \centering
            \caption{Pearson Correlation among Variables for Participants. p<0.05 (*), p<0.01 (**), p<0.001 (***).}

    \resizebox{\textwidth}{!}{
    \begin{tabular}{l|cccccccccc}
        \toprule
        \textbf{\#Projects} & 0.82*** & ~ & ~ & ~ & ~ & ~ & ~ \\ 
        \textbf{\#Interests} & 0.08*** & 0.08*** & ~ & ~ & ~ & ~ & ~ \\ 
        \textbf{\#Skills} & 0.15*** & 0.14*** & 0.24*** & ~ & ~ & ~ & ~ \\ 
        \textbf{Years of Experience} & 0.48*** & 0.48*** & 0.03*** & 0.14*** & ~ & ~ & ~ \\ 
        \textbf{Has Followers} & 0.26*** & 0.24*** & 0.12*** & 0.18*** & 0.23*** & ~ & ~ \\ 
        \textbf{Has Likes} & 0.3*** & 0.28*** & 0.11*** & 0.17*** & 0.28*** & 0.32*** & ~ \\ 
        \textbf{AVG. Weighted Winning} & 0.06*** & 0.05*** & 0.05*** & 0.04*** & 0.02*** & 0.06*** & 0.09*** \\ 
        \midrule
             & \textbf{\#Hackathons} & \textbf{\#Projects} & \textbf{\#Interests} & \textbf{\#Skills} & \textbf{Years of Experience} & \textbf{Has Followers} & \textbf{Has Likes} \\ 

        \bottomrule
        
    \end{tabular}}
    \Description{This table shows Pearson correlations among variables for hackathon participants. The correlations are as follows: Number of Projects correlates with Number of Hackathons (0.82***). Number of Interests correlates with Number of Hackathons (0.08***) and Number of Projects (0.08***). Number of Skills correlates with Number of Hackathons (0.15***), Number of Projects (0.14***), and Number of Interests (0.24***). Years of Experience correlates with Number of Hackathons (0.48***), Number of Projects (0.48***), Number of Interests (0.03***), and Number of Skills (0.14***). Has Followers correlates with Number of Hackathons (0.26***), Number of Projects (0.24***), Number of Interests (0.12***), Number of Skills (0.18***), and Years of Experience (0.23***). Has Likes correlates with Number of Hackathons (0.3***), Number of Projects (0.28***), Number of Interests (0.11***), Number of Skills (0.17***), Years of Experience (0.28***), and Has Followers (0.32***). Average Weighted Winning correlates with Number of Hackathons (0.06***), Number of Projects (0.05***), Number of Interests (0.05***), Number of Skills (0.04***), Years of Experience (0.02***), Has Followers (0.06***), and Has Likes (0.09***). All correlations are statistically significant, with p$<$0.001 (***) for all reported correlations.}
\label{tab:corr_creators}
\end{table*}

\begin{table*}[htbp]
    \centering
    
      \caption{Pearson Correlation among Variables for Collaborations. p<0.05 (*), p<0.01 (**), p<0.001 (***).}
    \resizebox{\textwidth}{!}{
    \begin{tabular}{l|ccccccccccc}
    \toprule
        \textbf{Collab. Proj. Repetition} & 0.49*** & ~ & ~ & ~ & ~ & ~ & ~ & ~ & ~ & ~ & ~ \\ 
        \textbf{Collab. Hack. Repetition} & 0.29*** & 0.86*** & ~ & ~ & ~ & ~ & ~ & ~ & ~ & ~ & ~ \\ 
        \textbf{Proj. Repetition} & 0.19*** & 0.48*** & 0.35*** & ~ & ~ & ~ & ~ & ~ & ~ & ~ & ~ \\ 
        \textbf{Hack. Repetition} & 0.14*** & 0.58*** & 0.76*** & 0.58*** & ~ & ~ & ~ & ~ & ~ & ~ & ~ \\ 
        \textbf{Interests} & 0.25*** & 0.13*** & 0.08*** & 0.29*** & 0.17*** & ~ & ~ & ~ & ~ & ~ & ~ \\ 
        \textbf{Common Interests} & -0.31*** & -0.01 & -0.01 & 0.09*** & 0.09*** & 0.42*** & ~ & ~ & ~ & ~ & ~ \\ 
        \textbf{Diff. Interests} & 0.46*** & 0.14*** & 0.09*** & 0.26*** & 0.14*** & 0.84*** & -0.14*** & ~ & ~ & ~ & ~ \\ 
        \textbf{Skills} & 0.6*** & 0.33*** & 0.23*** & 0.27*** & 0.18*** & 0.33*** & -0.07*** & 0.4*** & ~ & ~ & ~ \\ 
        \textbf{Common Skills} & -0.41*** & -0.04*** & -0.02*** & -0.01* & 0.02*** & 0 & 0.45*** & -0.27*** & 0.06*** & ~ & ~ \\ 
        \textbf{Diff. Skills} & 0.73*** & 0.34*** & 0.23*** & 0.25*** & 0.16*** & 0.31*** & -0.24*** & 0.48*** & 0.92*** & -0.33*** & ~ \\ 
        \textbf{Winner} & 0.09*** & 0.03*** & 0.05*** & 0.09*** & 0.1*** & 0.08*** & -0.02*** & 0.1*** & 0.12*** & -0.03*** & 0.13*** \\ 
        \midrule
         & \textbf{\#Participants} & \textbf{Collab. Proj. Repetition} & \textbf{Collab. Hack. Repetition} & \textbf{Proj. Repetition} & \textbf{Hack. Repetition} & \textbf{Interests} & \textbf{Common Interests} & \textbf{Diff. Interests} & \textbf{Skills} & \textbf{Common Skills} & \textbf{Diff. Skills} \\ 
        \bottomrule
    \end{tabular}}
    \Description{This table shows Pearson correlations among variables for collaborations at the project level. The correlations are as follows: Collaboration Project Repetition correlates with Number of Participants (0.49***). Collaboration Hackathon Repetition correlates with Number of Participants (0.29***) and Collaboration Project Repetition (0.86***). Project Repetition correlates with Number of Participants (0.19***), Collaboration Project Repetition (0.48***), and Collaboration Hackathon Repetition (0.35***). Hackathon Repetition correlates with Number of Participants (0.14***), Collaboration Project Repetition (0.58***), Collaboration Hackathon Repetition (0.76***), and Project Repetition (0.58***). Interests correlates with Number of Participants (0.25***), Collaboration Project Repetition (0.13***), Collaboration Hackathon Repetition (0.08***), Project Repetition (0.29***), and Hackathon Repetition (0.17***). Common Interests correlates with Number of Participants (-0.31***), Project Repetition (0.09***), Hackathon Repetition (0.09***), and Interests (0.42***). Different Interests correlates with Number of Participants (0.46***), Collaboration Project Repetition (0.14***), Collaboration Hackathon Repetition (0.09***), Project Repetition (0.26***), Hackathon Repetition (0.14***), Interests (0.84***), and Common Interests (-0.14***). Skills correlates with Number of Participants (0.6***), Collaboration Project Repetition (0.33***), Collaboration Hackathon Repetition (0.23***), Project Repetition (0.27***), Hackathon Repetition (0.18***), Interests (0.33***), Common Interests (-0.07***), and Different Interests (0.4***). Common Skills correlates with Number of Participants (-0.41***), Collaboration Project Repetition (-0.04***), Collaboration Hackathon Repetition (-0.02***), Project Repetition (-0.01*), Hackathon Repetition (0.02***), Common Interests (0.45***), Different Interests (-0.27***), and Skills (0.06***). Different Skills correlates with Number of Participants (0.73***), Collaboration Project Repetition (0.34***), Collaboration Hackathon Repetition (0.23***), Project Repetition (0.25***), Hackathon Repetition (0.16***), Interests (0.31***), Common Interests (-0.24***), Different Interests (0.48***), Skills (0.92***), and Common Skills (-0.33***). Winner correlates with Number of Participants (0.09***), Collaboration Project Repetition (0.03***), Collaboration Hackathon Repetition (0.05***), Project Repetition (0.09***), Hackathon Repetition (0.1***), Interests (0.08***), Common Interests (-0.02***), Different Interests (0.1***), Skills (0.12***), Common Skills (-0.03***), and Different Skills (0.13***). All correlations are statistically significant, with p$<$0.001 (**) for all reported correlations except for Project Repetition and Common Skills, which has p$<$0.05.}

    \label{tab:corr_collab}
\end{table*}

\begin{table*}[htbp]
    \centering
        \caption{Pearson Correlation among Variables in Hackathons. p<0.05 (*), p<0.01 (**), p<0.001 (***).}
      \resizebox{\textwidth}{!}{
    \begin{tabular}{l|cccccccccc}
    \toprule
    
        \textbf{\# act. Participants} & 0.92*** & ~ & ~ & ~ & ~ & ~ & ~ & ~ & ~ \\ 
        \textbf{\# Judges} & 0.11*** & 0.08*** & ~ & ~ & ~ & ~ & ~ & ~ & ~ \\ 
        \textbf{\# Sponsors} & 0.08*** & 0.08*** & 0.27*** & ~ & ~ & ~ & ~ & ~ & ~ \\ 
        \textbf{Onsite} & -0.17*** & -0.11*** & -0.08*** & -0.07*** & ~ & ~ & ~ & ~ & ~ \\ 
        \textbf{\# Days} & 0.04** & 0 & 0.05*** & -0.07*** & -0.34*** & ~ & ~ & ~ & ~ \\ 
        \textbf{\# Places} & 0.28*** & 0.21*** & 0.23*** & 0.23*** & 0.04** & -0.1*** & ~ & ~ & ~ \\ 
        \textbf{Competition (T)} & -0.22*** & -0.17*** & -0.07*** & 0.01 & 0.13*** & -0.07*** & 0.02 & ~ & ~ \\ 
        \textbf{Competition (I)} & -0.17*** & -0.14*** & -0.08*** & -0.03* & 0.05*** & 0 & -0.04** & 0.85*** & ~ \\ 
        \textbf{\# Winners} & 0.7*** & 0.62*** & 0.1*** & 0.14*** & -0.25*** & 0.05*** & 0.27*** & -0.18*** & -0.15*** \\ 
        \midrule
        & \textbf{\# sub. Projects} & \textbf{\# act. Participants} & \textbf{\# Judges} & \textbf{\# Sponsors} & \textbf{Onsite} & \textbf{\# Days} & \textbf{\# Places} & \textbf{Competition (T)} & \textbf{Competition (I)}  \\ 
    \bottomrule
    
    \end{tabular}}
    \Description{This table shows Pearson correlations among variables for hackathons. The correlations are as follows: Number of Active Participants correlates with Number of Submitted Projects (0.92***). Number of Judges correlates with Number of Submitted Projects (0.11***) and Number of Active Participants (0.08***). Number of Sponsors correlates with Number of Submitted Projects (0.08***), Number of Active Participants (0.08***), and Number of Judges (0.27***). Onsite correlates with Number of Submitted Projects (-0.17***), Number of Active Participants (-0.11***), Number of Judges (-0.08***), and Number of Sponsors (-0.07***). Number of Days correlates with Number of Submitted Projects (0.04**), Number of Judges (0.05***), Number of Sponsors (-0.07***), and Onsite (-0.34***). Number of Places correlates with Number of Submitted Projects (0.28***), Number of Active Participants (0.21***), Number of Judges (0.23***), Number of Sponsors (0.23***), Onsite (0.04**), and Number of Days (-0.1***). Competition (T) correlates with Number of Submitted Projects (-0.22***), Number of Active Participants (-0.17***), Number of Judges (-0.07***), Onsite (0.13***), and Number of Days (-0.07***). Competition (I) correlates with Number of Submitted Projects (-0.17***), Number of Active Participants (-0.14***), Number of Judges (-0.08***), Number of Sponsors (-0.03*), Onsite (0.05***), Number of Places (-0.04**), and Competition (T) (0.85***). Number of Winners correlates with Number of Submitted Projects (0.7***), Number of Active Participants (0.62***), Number of Judges (0.1***), Number of Sponsors (0.14***), Onsite (-0.25***), Number of Days (0.05***), Number of Places (0.27***), Competition (T) (-0.18***), and Competition (I) (-0.15***). Most correlations are statistically significant at p$<$0.001 (), with a few at p$<$0.01 () or p$<$0.05 (). Non-significant correlations are not reported.}
    \label{tab:corr_contexts}

\end{table*}

\section{Figures}

\begin{figure}[htbp]
    \centering
    \includegraphics[width=\linewidth]{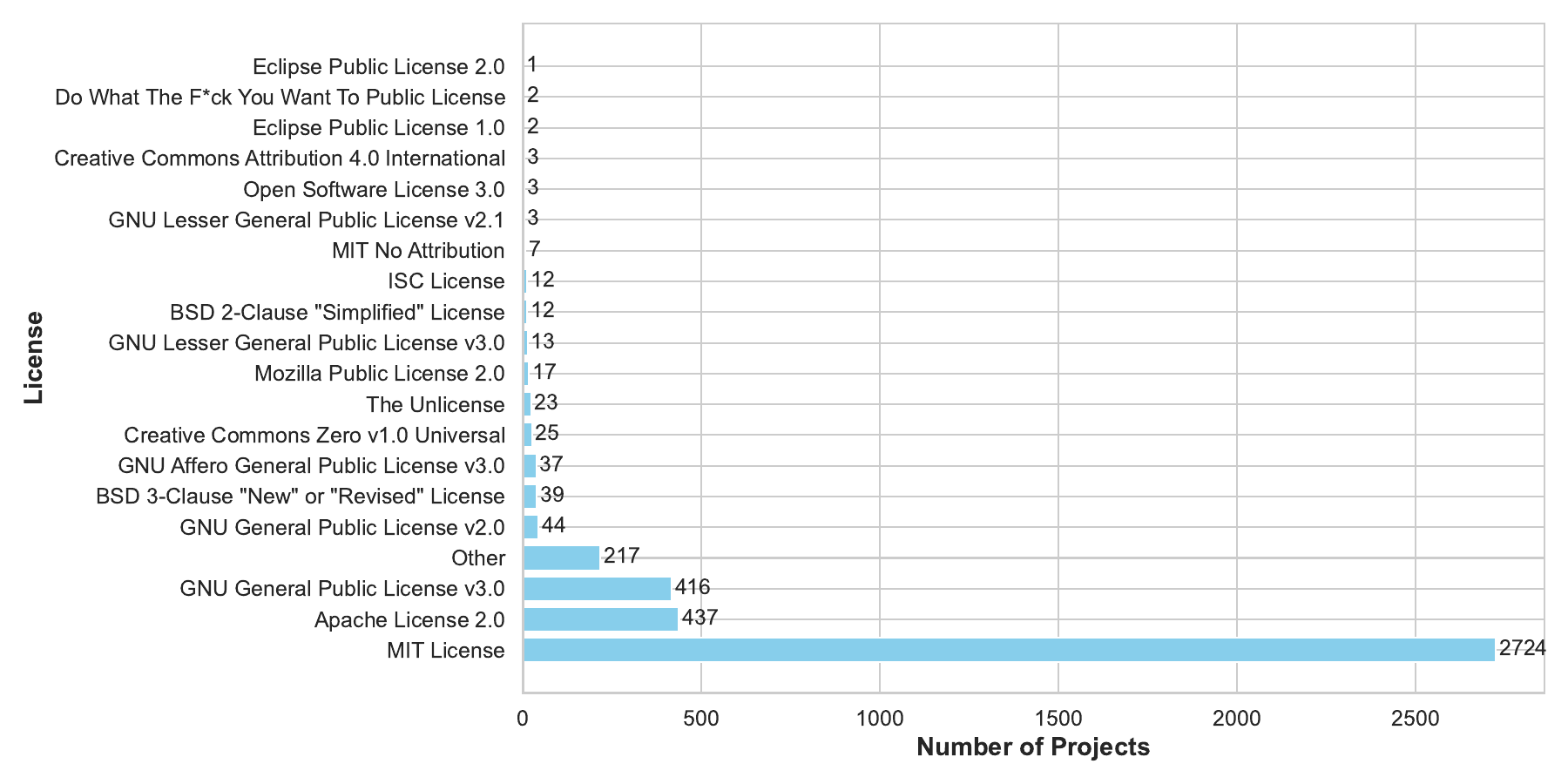}
    \caption{The Distribution of Licenses used in the Projects.}
    \label{fig:license}
    \Description{This figure represents the distribution of project licenses using a horizontal bar chart. The x-axis indicates the number of projects, and the y-axis lists the license types. The exact numbers of projects for each license type are as follows: MIT License (2,724), Apache License 2.0 (437), GNU General Public License v3.0 (416), Other (217), GNU General Public License v2.0 (44), BSD 3-Clause "New" or "Revised" License (39), GNU Affero General Public License v3.0 (37), Creative Commons Zero v1.0 Universal (25), The Unlicense (23), Mozilla Public License 2.0 (17), GNU Lesser General Public License v3.0 (13), BSD 2-Clause "Simplified" License (12), ISC License (12), MIT No Attribution (7), GNU Lesser General Public License v2.1 (3), Open Software License 3.0 (3), Creative Commons Attribution 4.0 International (3), Eclipse Public License 1.0 (2), Do What The F*ck You Want To Public License (2), and Eclipse Public License 2.0 (1). Each bar is labeled with these exact numbers for clarity. The chart uses light blue bars and bold axis labels to enhance readability and accessibility. }
\end{figure}

\begin{figure}[ht!]
    \centering
    \includegraphics[width=\linewidth]{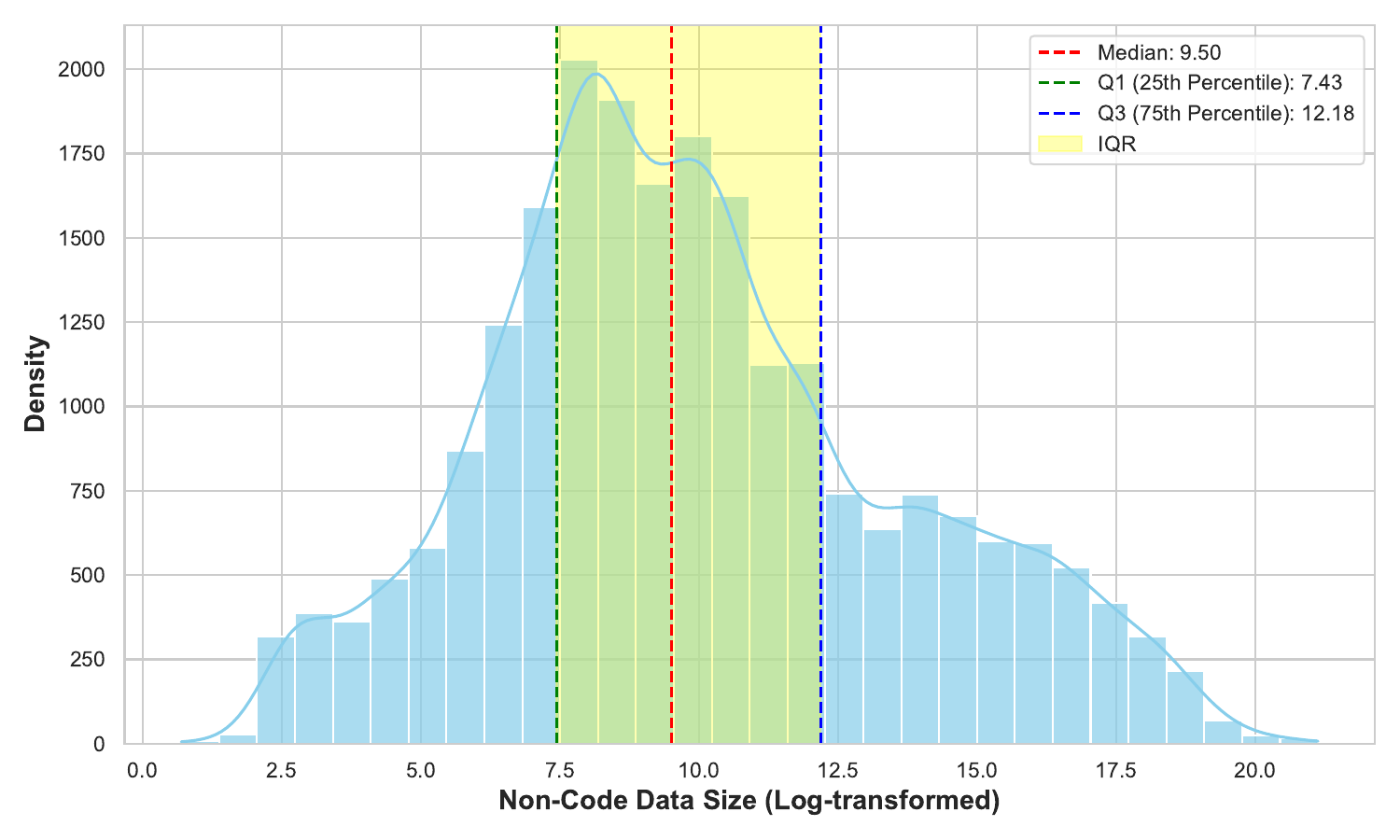}
    \caption{The Distribution of Non-Code Data Sizes in Projects with highlighted Interquartile Range (25\% to 75\%). 50\% of the data sizes are between 1807 to 268336 Bytes.}
    \label{fig:datasize_github_distribution}
    \Description{This figure visualizes the distribution of log-transformed non-code data sizes using a histogram with an overlaid kernel density estimation (KDE) curve to smooth the density representation. The x-axis represents the non-code data sizes after applying a logarithmic transformation, while the y-axis indicates the density of data points. Key statistical features are annotated, including the median, represented by a red dashed line, marking the 50th percentile where half of the data lies below and half above. The first quartile (Q1), shown as a green dashed line, represents the 25th percentile, below which 25\% of the data falls, and the third quartile (Q3), marked by a blue dashed line, denotes the 75th percentile, below which 75\% of the data lies. The interquartile range (IQR), highlighting the central 50\% of the data, is shaded in yellow between Q1 and Q3, providing a visual summary of the spread of the middle portion of the dataset. This histogram effectively showcases the central tendency and variability of non-code data sizes, with distinct color contrasts and annotation styles to enhance accessibility for users with visual impairments.}
\end{figure}

\begin{figure*}[htbp]
    \centering
    \includegraphics[width=\linewidth]{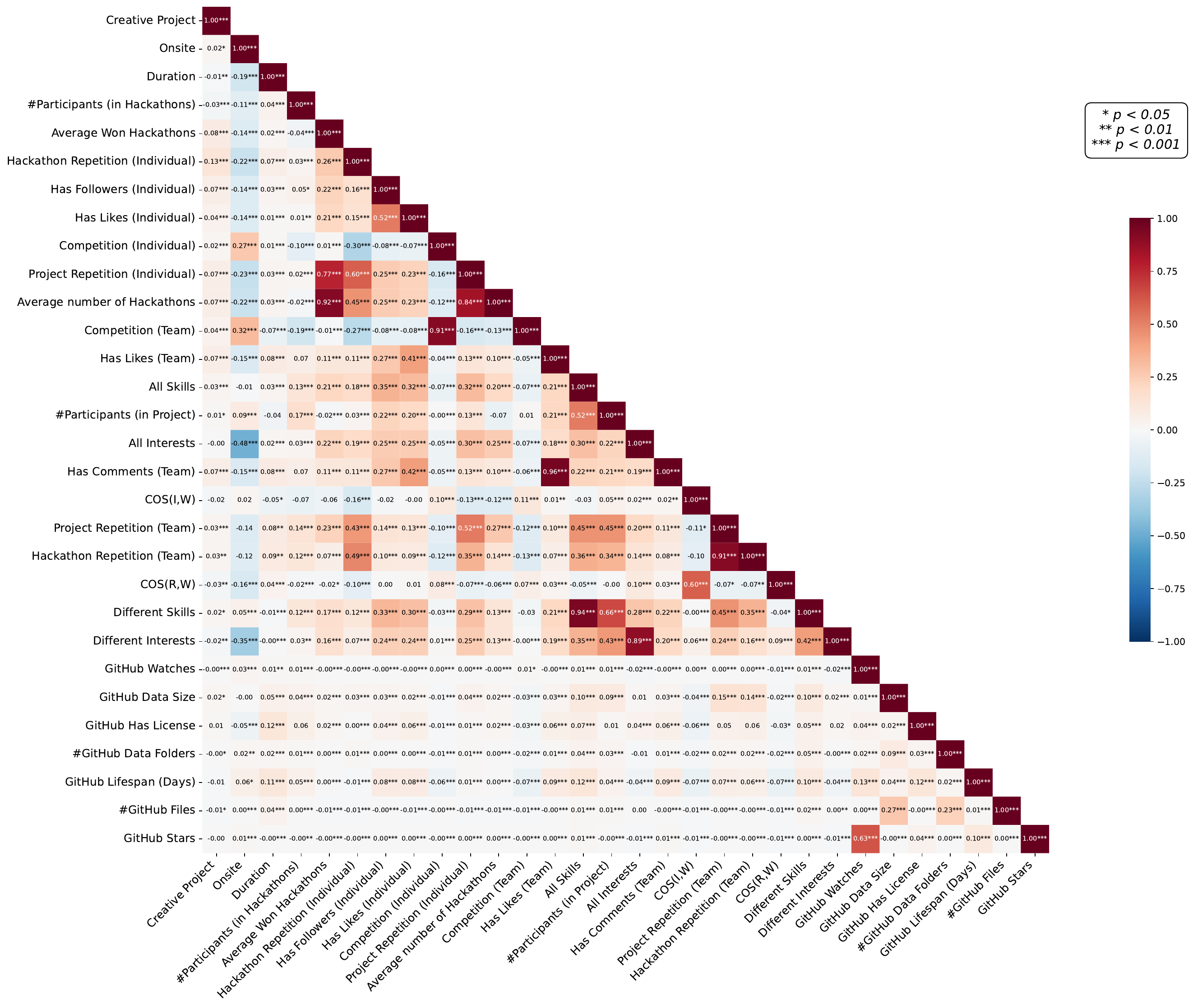}
    \caption{Spearman's Rank Correlation Coefficients among Variables.}
    \label{fig:correlation_mixed_effects}
    \Description{The heatmap illustrates the spearmann's rank correlation coefficients among variables. The strongest positive correlations are observed between team competition and individual competition (r = 0.91, p < 0.001), and between average number of hackathons and hackathon repetition at the individual level (r = 0.92, p < 0.001). Project participation shows moderate positive correlations with both skills and interests (r = 0.52, p < 0.001). GitHub metrics demonstrate weaker but significant correlations: GitHub lifespan correlates with project characteristics (r = 0.12, p < 0.001), while GitHub stars show minimal correlation with most variables (r < 0.10). Notably, onsite participation shows negative correlations with several variables, including hackathon repetition (r = -0.22, p < 0.001) and has followers (r = -0.14, p < 0.001). The data also reveals that creative projects have weak to moderate positive correlations with individual engagement metrics, such as hackathon repetition (r = 0.13, p < 0.001) and project repetition (r = 0.07, p < 0.001). Duration shows minimal correlation with most variables, with the strongest being a negative correlation with onsite participation (r = -0.19, p < 0.001).}
\end{figure*}

\clearpage
\section{Algorithm}\label{appendix:algorithms}

\begin{algorithm}
\SetAlgoLined
\KwIn{Dataset containing projects, hackathons, and participants}
\KwOut{creator2creator-project, creator2creator-hackathon, collaboration repetition and repetition metrics}

\textbf{Initialize:} \\
$creator2creator\_project \gets \emptyset$ \\
$creator2creator\_hackathon \gets \emptyset$ \\
$collaboration\_repetition\_project \gets \emptyset$ \\
$collaboration\_repetition\_hackathon \gets \emptyset$ \\
$repetition\_project \gets \emptyset$ \\
$repetition\_hackathon \gets \emptyset$ \\

\ForEach{project $p$ in dataset['projects']}{
    creators $C_p \gets p['creators']$ \\
    \ForEach{pair of creators $(c_i, c_j)$ in $C_p$}{
        \If{$(c_i, c_j)$ or $(c_j, c_i)$ exists in $creator2creator\_project$}{
            $creator2creator\_project[(c_i, c_j)] \gets creator2creator\_project[(c_i, c_j)] + 1$
        }
        \Else{
            $creator2creator\_project[(c_i, c_j)] \gets 1$
        }
    }
    \ForEach{creator $c$ in $C_p$}{
        $collaboration\_repetition\_project[c] \gets collaboration\_repetition\_project[c] + 1$ \\
        $repetition\_project[c] \gets repetition\_project[c] + collaboration\_repetition\_project[c]$
    }
}

\ForEach{hackathon $h$ in dataset['hackathons']}{
    creators $C_h \gets h['creators']$ \\
    \ForEach{pair of creators $(c_i, c_j)$ in $C_h$}{
        \If{$(c_i, c_j)$ or $(c_j, c_i)$ exists in $creator2creator\_hackathon$}{
            $creator2creator\_hackathon[(c_i, c_j)] \gets creator2creator\_hackathon[(c_i, c_j)] + 1$
        }
        \Else{
            $creator2creator\_hackathon[(c_i, c_j)] \gets 1$
        }
    }
    \ForEach{creator $c$ in $C_h$}{
        $collaboration\_repetition\_hackathon[c] \gets collaboration\_repetition\_hackathon[c] + 1$ \\
        $repetition\_hackathon[c] \gets repetition\_hackathon[c] + collaboration\_repetition\_hackathon[c]$
    }
}

\Return $collaboration\_repetition\_project$, $collaboration\_repetition\_hackathon$, \\
$repetition\_project$, $repetition\_hackathon$

\caption{Project and Hackathon Collaboration Algorithm}
\Description{This algorithm processes a dataset containing projects, hackathons, and participants to calculate collaboration metrics. It initializes six empty dictionaries: creator2creator_project, creator2creator_hackathon, collaboration_repetition_project, collaboration_repetition_hackathon, repetition_project, and repetition_hackathon. The algorithm iterates through each project in the dataset, identifying pairs of creators and updating the creator2creator_project dictionary with the number of times each pair collaborates. It also updates collaboration_repetition_project and repetition_project for each creator, counting their collaborations and cumulative collaborations. The algorithm then performs a similar process for hackathons, updating creator2creator_hackathon, collaboration_repetition_hackathon, and repetition_hackathon. Finally, it returns the collaboration repetition and repetition metrics for both projects and hackathons. This algorithm helps quantify the frequency and patterns of collaborations among creators in both project and hackathon contexts.}
\label{algo:collab_interaction}
\end{algorithm}

%TC:endignore

\end{document}